\documentclass[prd,12pt]{article}

\usepackage{diagbox}
\usepackage{tikz}
\usepackage{amsmath,amssymb,graphicx,multirow,xspace,slashed,array,booktabs}
\usepackage[colorlinks=true,urlcolor=blue,anchorcolor=blue,citecolor=blue,filecolor=blue,linkcolor=blue,menucolor=blue,pagecolor=blue]{hyperref}
\usepackage[compress,numbers]{natbib}
\usepackage{placeins}
\usepackage{subcaption}
\usepackage{booktabs}
\usepackage{blindtext, rotating}
\usepackage{afterpage}
\usepackage{enumitem}
\usepackage{ marvosym }
\usepackage{authblk} 
\usepackage{verbatim}
\usepackage{soul} 
\usepackage[normalem]{ulem}
\usepackage{pifont}
\usepackage{booktabs}
\usepackage{bm}
\usepackage{cleveref}
\usepackage{siunitx}
\usepackage{tikz}
\usepackage{tikz-feynman}
\tikzfeynmanset{compat=1.1.0}
\usepackage{cases}
\usepackage{cancel}

\usepackage{floatrow}
\newfloatcommand{capbtabbox}{table}[][\FBwidth]

\usepackage[font=footnotesize,labelfont=bf]{caption}

\usepackage{lineno}

\allowdisplaybreaks

\long\def\symbolfootnote[#1]#2{\begingroup%
\def\thefootnote{\fnsymbol{footnote}}\footnote[#1]{#2}\endgroup}

\allowdisplaybreaks

\makeatletter
	
	\@addtoreset{equation}{section}
\makeatother

\newcommand{\newc}{\newcommand}
\newc{\gsim}{\lower.7ex\hbox{$\;\stackrel{\textstyle>}{\sim}\;$}}
\newc{\lsim}{\lower.7ex\hbox{$\;\stackrel{\textstyle<}{\sim}\;$}}
\newc{\gev}{\,{\rm GeV}}
\newc{\mev}{\,{\rm MeV}}
\newc{\ev}{\,{\rm eV}}
\newc{\kev}{\,{\rm keV}}
\newc{\tev}{\,{\rm TeV}}

\def\ln{\mathop{\rm ln}}
\def\tr{\mathop{\rm tr}}
\def\Tr{\mathop{\rm Tr}}

\def\Re{\mathop{\rm Re}}

\newc{\mz}{M_Z}
\newc{\mpl}{M_*}
\newc{\mw}{m_{\rm weak}}
\newc{\nr}[1]{N^c_R{}_{#1}}

\newcommand{\dd}{\mathrm{d}}
\newcommand{\Mpl}{M_{\rm Pl}}

\newcommand{\abs}[1]{\left\vert {#1} \right\vert}

\usepackage{accents}
\newlength{\dhatheight}

\def\beq{\begin{equation}}
\def\eeq{\end{equation}}
\newcommand{\bea}{\begin{eqnarray}\begin{aligned}}
\newcommand{\eea}{\end{aligned}\end{eqnarray}}
\def\bitem{\begin{itemize}}
\def\eitem{\end{itemize}}

\newcommand{\lmk}{\left(}  
\newcommand{\rmk}{\right)}
\newcommand{\lkk}{\left[}  
\newcommand{\rkk}{\right]}

\newcommand{\del}{\partial}

\newcommand{\half}{\frac{1}{2}}
\newcommand{\vev}[1]{ \left\langle {#1} \right\rangle }

\newcommand{\1}{\mbox{1}\hspace{-0.25em}\mbox{l}}

\begin{document}
\baselineskip 0.6cm

\begin{titlepage}

\vspace*{-0.5cm}

\thispagestyle{empty}

\begin{center}

\vskip 1cm

{\LARGE
Deconstructing the Extra-Dimensional Axion
}

\vskip 1.5cm
{Shihwen Hor$^{1,2}$, Yuichiro Nakai$^{1,2}$, Motoo Suzuki$^{3,4,5}$, and Junxuan Xu$^{1,2}$}
\\*[10pt]
$^1${\it \normalsize Tsung-Dao Lee Institute, Shanghai Jiao Tong University, \\
No.~1 Lisuo Road, Pudong New Area, Shanghai, 201210, China} \\*[3pt]
$^2${\it \normalsize School of Physics and Astronomy, Shanghai Jiao Tong University, \\
800 Dongchuan Road, Shanghai, 200240, China} \\*[3pt]
$^3${\it \normalsize SISSA International School for Advanced Studies, \\
Via Bonomea 265, 34136, Trieste, Italy} \\
$^4${\it \normalsize INFN, Sezione di Trieste, Via Valerio 2, 34127, Italy}\\
$^5${\it 
\normalsize IFPU, Via Beirut 2, 34014, Trieste, Italy}

\vskip 1.0cm

\end{center}

\begin{abstract}

We present a four-dimensional deconstruction of the extra-dimensional axion arising from a $U(1)$ gauge theory in a five-dimensional orbifold, where the axion is identified with the Wilson line of the $U(1)$ gauge field and its coupling to QCD is generated by a 5D Chern-Simons (CS) term. We construct the corresponding 4D moose (quiver) gauge theory with link scalar fields, in which the axion emerges as a collective pseudo-Nambu-Goldstone boson. The axion-gluon coupling is described by a gauged Wess-Zumino-Witten term, providing the 4D counterpart of the 5D CS term. We further analyze non-perturbative effects from zero-mode and ``fractional'' instanton configurations. While the latter is exponentially suppressed in the regime corresponding to the 5D description, ensuring consistency with the higher-dimensional picture, we point out that this suppression can break down for smaller instantons whose inverse size exceeds the 5D cutoff scale, leading to a potentially significant effect. We also study axion potentials induced by bulk matter fields and boundary-localized symmetry-breaking operators, reproducing the characteristic nonlocal suppression associated with propagation in the extra dimension. Our construction provides a renormalizable 4D framework with a transparent understanding of the axion shift symmetry and its quality.

\end{abstract}

\flushbottom

\end{titlepage}

\section{Introduction
\label{sec:introduction}}

The strong CP problem remains a compelling motivation for physics beyond the Standard Model (SM). The most popular approach to the problem is provided by the Peccei-Quinn (PQ) mechanism
\cite{Peccei:1977hh}, in which a pseudo-Nambu-Goldstone Boson (pNGB) associated with spontaneous breaking of a global $U(1)_{\rm PQ}$ symmetry, the axion
\cite{Weinberg:1977ma,Wilczek:1977pj}, dynamically relaxes the CP-violating $\theta$ parameter to zero at the minimum of its potential. In conventional four-dimensional realizations, however, the PQ symmetry is typically imposed by hand and is vulnerable to explicit breaking by ultraviolet (UV) effects, leading to the well-known axion quality problem
\cite{Dine:1986bg,Barr:1992qq,Kamionkowski:1992mf,Kamionkowski:1992ax,Holman:1992us,Kallosh:1995hi,Carpenter:2009zs}.\footnote{There are various attempts to address the axion quality problem such as composite axion models
\cite{Kim:1984pt,Choi:1985cb,Randall:1992ut,Izawa:2002qk,Yamada:2015waa,Redi:2016esr,DiLuzio:2017tjx,Lillard:2017cwx,Lillard:2018fdt,Gavela:2018paw,Lee:2018yak,Yamada:2021uze,Ishida:2021avk,Contino:2021ayn,Sato:2025rok,Hor:2026dlb}, conformal axion models
\cite{Nakai:2021nyf,Nakagawa:2023shi,Nakagawa:2024kcb}, warped extra dimension models
\cite{Flacke:2006ad,Cox:2019rro,Bonnefoy:2020llz,Lee:2021slp,Csaki:2026qjl}, visible axion models
\cite{Rubakov:1997vp,Berezhiani:2000gh,Hook:2014cda,Fukuda:2015ana,Gherghetta:2016fhp,Dimopoulos:2016lvn,Gherghetta:2020ofz,Alves:2017avw,Liu:2021wap,Girmohanta:2024nyf,Iwai:2026pgz} and gauge symmetry protection models
\cite{Cheng:2001ys,Harigaya:2013vja,Fukuda:2017ylt,Fukuda:2018oco,Ibe:2018hir,Choi:2020vgb}.}

{\it Extra-dimensional realizations of the axion} provide a geometrical origin of the PQ symmetry and offer a promising route toward solving the axion quality problem
\cite{Choi:2003wr,Reece:2025thc,Craig:2024dnl,Choi:2026kxu}. In a five-dimensional spacetime with a compact extra dimension, the axion can be identified with the Wilson line of a bulk $U(1)$ gauge field along the extra dimension. The axion-gluon coupling is naturally introduced by a Chern-Simons term in the bulk. In this framework, the axion shift symmetry originates from higher-dimensional gauge invariance and is therefore protected against local UV corrections. 
Despite such a beautiful feature of the extra-dimensional axion, a higher-dimensional theory is inherently non-renormalizable and should be regarded as an effective description valid only below a cutoff scale.
It is then unclear to what extent the high-quality nature of the extra-dimensional axion is maintained under a proper UV completion of the theory, which is the central question underlying the present study.

One possible UV completion of the extra-dimensional axion is provided by a four-dimensional renormalizable theory realized through {\it dimensional deconstruction}
\cite{Arkani-Hamed:2001kyx,Hill:2000mu,Arkani-Hamed:2001nha,Cheng:2001vd,Arkani-Hamed:2003xts}, in which a higher-dimensional gauge theory emerges in the infrared from a product of gauge groups connected by bifundamental link scalar fields. When the link fields acquire vacuum expectation values (VEVs), the product gauge groups are Higgsed to a diagonal subgroup, and the resulting spectrum contains a tower of massive vector bosons that closely mimics Kaluza-Klein (KK) excitations of the compact extra dimension. In the limit of many sites with appropriately scaled parameters, the discrete gauge moose structure approaches a continuous fifth dimension, reproducing higher-dimensional locality and gauge dynamics. At the same time, since the theory is fundamentally four-dimensional, it remains renormalizable and provides a transparent setting in which even non-perturbative effects can be estimated with confidence. Therefore, dimensional deconstruction offers a natural framework to revisit the extra-dimensional axion and to test whether its geometrical solution to the axion quality problem can survive in a genuine four-dimensional completion.

While the authors of Ref.~\cite{Hill:2002kq} initiated the study of the Wilson-line pNGB arising in a latticized 5D $U(1)$ gauge theory as a realization of the high-quality QCD axion, the analysis focused on the case of $S^1$ compactification,
which is phenomenologically unrealistic as a massless $U(1)$ gauge field remains in the spectrum. Moreover, the discussion on the effect of a bulk matter field on the axion potential is qualitatively different from that of the more realistic orbifold case: on an orbifold, the Wilson-line dependence of a bulk matter field can be completely removed by a gauge transformation, which is impossible on $S^1$, and an axion potential can be generated only in the presence of explicit breaking sources on the boundaries.

In this paper, we construct a deconstructed version of the orbifold extra-dimensional axion and systematically study its properties. We show that the axion arises as a collective pNGB associated with a global shift symmetry acting on the link fields, thereby reproducing the Wilson-line nature of the extra-dimensional axion. 
The axion-gluon coupling is encoded in a gauged Wess-Zumino-Witten term, which provides a precise four-dimensional counterpart of the 5D Chern-Simons term. 
This extends the correspondence between 5D Chern-Simons terms and gauged WZW terms discussed in Refs.~\cite{Hill:2004uc,Skiba:2002nx}, where only the $SU(N)^3$ case was considered, to the mixed $U(1)-SU(3)^2$ case relevant for the axion-gluon coupling. 
We further investigate non-perturbative effects and the resulting axion potential in the 4D deconstructed theory.
Previous studies of deconstructed 5D instanton effects focused on the zero-mode instanton contribution and regarded the ``fractional'' instanton configurations associated with individual gauge sites as lattice artifacts~\cite{Poppitz:2002ac,Gherghetta:2020keg}. 
We first work in the instanton-size regime relevant to the 5D description, explicitly compute the zero-mode instanton contribution, and show that the ``fractional'' instanton contributions are exponentially suppressed, thereby ensuring consistency with the higher-dimensional picture.
We then extend the analysis to the smaller-instanton regime intrinsic to the deconstructed theory, where the inverse instanton size exceeds the 5D cutoff scale.
In this region, the exponential suppression factor becomes ineffective, suggesting sizable instanton contributions beyond the estimate obtained from the 5D regime alone.
Finally, we analyze the effects of bulk matter fields and boundary-localized symmetry-breaking operators, demonstrating that the resulting axion potential exhibits the characteristic nonlocal suppression expected from propagation along the extra dimension.

The rest of the paper is organized as follows. Sec.~\ref{sec:review} starts with a review of the extra-dimensional axion in a five-dimensional orbifold setup. Then, in Sec.~\ref{sec:deconstruction}, we construct its deconstructed counterpart and establish the correspondence between the two descriptions. In Sec.~\ref{sec:WZW}, we derive the axion-gluon coupling from a gauged Wess-Zumino-Witten term. Sec.~\ref{sec:quality} analyzes the axion potential and discusses implications for the axion quality problem. Sec.~\ref{sec:Discussion} is devoted to conclusions and discussions.
Some details are summarized in appendices.

\section{Review of the Extra-Dimensional Axion
\label{sec:review}}
We consider a $U(1)\times SU(3)$ gauge theory in a flat 5D spacetime with the fifth dimension compactified on the orbifold $S^1/Z_2$ as the minimal setup of the extra-dimensional axion, following the discussion in Ref.~\cite{Choi:2003wr}. 
The gauge-sector action is
\begin{align}
S_{5}^{\rm gauge} &= \int \dd^5 x \lkk
-\frac{1}{4g_5^2} \mathcal{F}_{MN} \mathcal{F}^{MN}
-\frac{1}{2g_{5,c}^2}\tr \lmk \mathcal{G}_{MN}\mathcal{G}^{MN} \rmk\rkk +S_5^{\rm CS} + S_5^{\rm GF} \, , \label{5D_gauge}
\end{align}
with
\begin{align}
S_5^{\rm CS}&=-\int \dd^5 x \, \frac{\kappa_{\rm CS}}{32\pi^2}\epsilon^{MNPQR}A_M\tr \lmk \mathcal{G}_{NP}\mathcal{G}_{QR}\rmk\, ,\label{5D_CS}\\
S_5^{\rm GF} &= \int \dd^5 x \lkk -\frac{1}{2\xi_1 g_5^2}\lmk\partial^\mu A_\mu-\xi_1\partial_y A_5\rmk^2 - \frac{1}{2\xi_c g_{5,c}^2}\lmk\partial^\mu G^a_\mu-\xi_c\partial_y G^a_5\rmk^2 \rkk\notag\\
 & - \int d^4x\left[\frac{1}{2 \xi_{1,\rm b} g_5^2}\bigl(\partial^\mu A_\mu\mp \xi_{1,\rm b}A_5\bigr)^2
 +\frac{1}{2 \xi_{c,\rm b} g_{5,c}^2}\bigl(\partial^\mu G^a_\mu\mp \xi_{c, \rm b}G_5^a\bigr)^2\right]_{0,\pi R}\label{5D_gauge_fixing}\, ,
\end{align}
where we take the metric $\eta_{MN}={\rm diag}(+,-,-,-,-)$ with $M=0,1,2,3,5$ and $x^5\equiv y\in [0,\pi R]$,
$A_M, G_M\equiv G_M^a T^a$ ($a=1,\cdots,8$) represent the $U(1)$ and $SU(3)$ gauge fields, respectively, with $T^a$ the $SU(3)$ generators normalized as $\tr\lmk T^a T^b\rmk=\frac{1}{2}\delta^{ab}$, and $\mathcal{F}_{MN}=\del_MA_N-\del_NA_M,~ \mathcal{G}_{MN}=\del_MG_N-\del_NG_M-i[G_M,G_N] $ are the corresponding field strengths.
The gauge fields $A_M, G_M$ have mass dimension $1$ as in the 4D theory while the corresponding gauge couplings $g_5,~g_{5,c}$ have mass dimension $-\half$.
Therefore the 5D gauge theory is non-renormalizable and should be considered as an effective theory with a cutoff.
Eq.~\eqref{5D_CS} gives the expression of the 5D Chern-Simons (CS) term, where $\epsilon^{MNPQR}$ is the 5D totally anti-symmetric tensor with $\epsilon^{01235}=+1$ and the coefficient $\kappa_{\rm CS}$ is a dimensionless parameter and taken to be an integer.
Eq.~\eqref{5D_gauge_fixing} shows the gauge fixing terms which cancel the mixing between $A_\mu(G_\mu)$ and $A_5(G_5)$ components both in the bulk and on the boundaries\footnote{In general, the gauge fixing parameters on the two boundaries are different. We take the common $\xi_{\rm b}$ just for simplicity.} when we apply the action principle to derive the equations of motion and the boundary conditions. The second line in Eq.~\eqref{5D_gauge_fixing} sums over the two boundary gauge-fixing terms where the $-$ sign is for $y=0$ and $+$ for $y=\pi R$.

Here we impose different types of boundary conditions on the $A_M$ and $G_M$: Dirichlet boundary conditions for the $A_\mu \, (G_5)$ components\footnote{Dirichlet boundary conditions require the vanishing of the field variations, which is equivalent with that the field values are taken constant. Due to the 4D Lorentz symmetry, $A_\mu$ has to vanish at the boundaries while $G_5$ in general can have a non-zero value. We take $G_5|_{0,\pi R}=0$ for simplicity.} and Neumann boundary conditions for $A_5 \, (G_\mu)$,
\begin{align}
A_\mu|_{0,\pi R}=0\, ,\qquad \partial_y A_5|_{0,\pi R}=0\, ,\label{5D_BC_u1}\\[1ex]
\partial_y G_\mu|_{0,\pi R}=0\, ,\qquad G_5|_{0,\pi R}=0\, .\label{5D_BC_su3}
\end{align}
As a result, the (infinitesimal) gauge transformation parameters $\alpha_1(x,y), \, \alpha_3(x,y)\equiv \alpha_3^a(x,y) T^a$, which are defined by
\begin{align}
    A_M(x,y)&\rightarrow A_M(x,y) - \del_M\alpha_1(x,y)\, , \label{gauge_transform_u1}\\[1ex]
    G_M(x,y) &\rightarrow G_M(x,y) - \del_M\alpha_3(x,y) + i[\alpha_3(x,y),G_M(x,y)]\, ,
\end{align}
are also constrained by the boundary conditions,
\begin{align}
    \del_\mu \alpha_1(x,y=0,\pi R)=0\, ,\label{u1_parambc}\\
    \del_y\alpha_3(x,y=0,\pi R)=0\, , \label{su3_parambc}
\end{align}
where the constraint on $\alpha_1$ from the $A_5$ component is given by the second derivative in $y$ and thus not relevant while the constraint on $\alpha_3$ from the $G_\mu$ component is automatically satisfied as long as Eq.~\eqref{su3_parambc} is satisfied.
Then, the $U(1)$ gauge symmetry is reduced to a global symmetry on the boundaries and the remaining bulk gauge symmetry is parameterized by $\alpha^{\rm re}_1(x,y)$ that satisfies
\begin{align}
    \alpha_1^{\rm re}(x,0) = \alpha_1^{\rm re}(x,\pi R) = 0\, .\label{remaining_u1}
\end{align}
In contrast, the bulk $SU(3)$ gauge symmetry remains even on the boundaries.

One might worry that the CS term~\eqref{5D_CS} seems not gauge-invariant due to the existence of the boundaries.
We can see that it varies under the $U(1)$ gauge transformation~\eqref{gauge_transform_u1} as
\begin{align}
	\delta S^{\rm CS}_5 &= \int\dd^4 x\int_0^{\pi R}\dd y \, \frac{\kappa_{\rm CS}}{32\pi^2}\epsilon^{MNPQR}\del_M \alpha_1^{\rm re}(x,y) \tr \lmk \mathcal{G}_{NP}\mathcal{G}_{QR}\rmk \notag\\
	&= \frac{\kappa_{\rm CS}}{32\pi^2} \int\dd^4 x\int_0^{\pi R}\dd y \, \epsilon^{\mu\nu\rho\sigma}\del_5 \lkk \alpha_1^{\rm re}(x,y) \tr \lmk \mathcal{G}_{\mu\nu}\mathcal{G}_{\rho\sigma}\rmk \rkk + \text{4D total derivatives}  \notag \\
	&= \kappa_{\rm CS}\int\dd^4 x \lkk\alpha^{\rm re}_1(x,y=\pi R) \mathcal{A}(y=\pi R)- \alpha_1^{\rm re}(x,y=0) \mathcal{A}(y=0) \rkk\, , \label{delta_CS}
\end{align}
where the 4D total derivatives in the second line vanish after the integration over 4D spacetime and the quantity,
\begin{align}
	\mathcal{A}=\frac{1}{16\pi^2}  \tr \lmk \mathcal{G}_{\mu\nu}\widetilde{\mathcal{G}}^{\mu\nu}\rmk\, ,
\end{align}
is the usual 4D chiral anomaly. 
Here $\widetilde{\mathcal{G}}^{\mu\nu}\equiv \frac{1}{2}\epsilon^{\mu\nu\rho\sigma}\mathcal{G}_{\rho\sigma}$ is the dual field strength with the 4D totally anti-symmetric tensor defined as $\epsilon^{\mu\nu\rho\sigma} = \epsilon^{5\mu\nu\rho\sigma}$. 
The variation~\eqref{delta_CS} turns out to be zero because of the constraint~\eqref{remaining_u1} on $\alpha_1^{\rm re}(x,y)$.
Later we will see that the 5D CS term provides the axion-gluon coupling.

Performing the Kaluza-Klein (KK) decompositions,
\begin{align}
	A_\mu(x,y) = \sum_{n=0}^{\infty}A_{\mu,n}(x)f^A_{n}(y)\, , &\quad A_5(x,y) = \sum_{n=0}^{\infty}A_{5,n}(x)g^A_{n}(y)\, , \label{KK_expansion_u1}\\
	G_\mu(x,y) = \sum_{n=0}^{\infty}G_{\mu,n}(x)f^G_{n}(y)\, ,&\quad G_5(x,y) = \sum_{n=0}^{\infty}G_{5,n}(x)g^G_{n}(y)\, ,\label{KK_expansion_su3}
\end{align}
we find the 4D effective theory of a tower of KK modes whose 5D profiles are given by
\begin{align}
	f_n^A=g_n^G &= \sqrt{2} \sin\frac{ny}{R}  \quad \text{for} ~ n\geq 0\, ,\\[1ex]
	g_0^A=f_0^G=1 \, &, \quad g_n^A=f_n^G = \sqrt{2} \cos\frac{ny}{R}  \quad \text{for} ~ n\geq1\, ,
\end{align}
with the KK masses $m_n = \frac{n}{R}$.
The detailed derivation of the 5D EoMs, general boundary conditions and KK spectrum is given in Appendix~\ref{sec:appendix_5D_spectrum}. 
The 4D effective gauge couplings can be read off from the kinetic terms, for example, in the $U(1)$ case,
\begin{align}
    S_5&\supset \int\dd^4 x\int\dd y\lmk-\frac{1}{4g_5^2}\mathcal{F}_{\mu\nu}(x,y)\mathcal{F}^{\mu\nu}(x,y)\rmk \notag\\
    &= \int\dd^4 x\sum_{m,n=0}^\infty -\frac{1}{4g_5^2}\lmk \int\dd y \, f_m^A(y) f_n^A(y)\rmk \mathcal{F}_{\mu\nu,m}(x)\mathcal{F}^{\mu\nu}_n(x)\notag \\ 
    &=\int\dd^4x \lmk - \frac{1}{4 g_{4}^2}\sum_{n=1}^\infty \mathcal{F}_{\mu\nu,n}(x)\mathcal{F}^{\mu\nu}_n(x) \rmk\, , 
\end{align}
where the orthogonality of $f_n^A$ is used.
The 4D effective coupling $g_4$ is defined as
\begin{align}
    \frac{1}{g_4^2} = \frac{1}{g_5^2}\lmk \int_0^{\pi R}\dd y \, f_{n\neq 0}^A(y) f_{n\neq 0}^A(y)\rmk = \frac{\pi R}{g_5^2} \, .
\end{align}
The similar discussion is also applied to the $SU(3)$ case. Thus, we obtain
\begin{align}
    \frac{1}{g_{4,c}^2} = \frac{\pi R}{g_{5,c}^2} \, .
\end{align}
To summarize, the KK spectrum is therefore
\begin{align}
 &\text{One scalar zero mode:} ~~~ A_{5,0}(x)=\frac{1}{\pi R}\int \dd y ~A_5(x,y)\, ,\label{KKspec_A50}\\
 &\text{One gauge field zero mode:} ~~~ G_{\mu,0}^a(x)=\frac{1}{\pi R} \int \dd y ~G_{\mu}^a(x,y) \, ,\label{KKspec_Gmu0}\\[4pt]
 &\text{Towers of massive gauge bosons with masses $m_{V,n}={n}/{R}\,$:}  \label{KKspec_AG}\\
 & ~~~ V_{\mu,n}(x)= \frac{1}{\pi R}\int\dd y ~f_n^V(y) V_\mu(x,y)\, ,~~~ V_{5,n}(x)=\frac{1}{\pi R}\int\dd y ~g_n^V(y) V_5(x,y)\, ,~~~ n\geq 1\, . \notag
\end{align}
Here, the fifth component $V_{5,n\geq 1}$ plays the same role as the Goldstone boson in the Higgs mechanism.
At energies below the compactification scale $R^{-1}$, the only surviving degrees of freedom are the zero modes $A_{5,0}$ and $G_{\mu,0}^a$. The former is identified as the axion degree of freedom and the latter is the gluon field.
One common definition for the axion is given by the phase of the Wilson line, 
\beq \theta_5(x) \equiv \int_0^{\pi R}\dd y \, A_5(x,y)=A_{5,0}(x) \pi R\, ,\eeq
where $\theta_5(x)$ is a dimensionless field and all the higher KK modes vanish in the integral. The decay constant can be read off from the kinetic term of the $U(1)$ gauge field in Eq.~\eqref{5D_gauge},
\beq \mathcal{L}_{\rm kin}=\lmk\int_0^{\pi R}\dd y\rmk \frac{1}{2g_5^2}\del_\mu A_{5,0}\del^\mu A_{5,0}=\frac{1}{2g_5^2\pi R}\del_\mu\theta_5 \del^\mu\theta_5\equiv  \frac{1}{2}f_a^2\del_\mu\theta_5 \del^\mu\theta_5\, ,\eeq
with $f_a^2=\frac{1}{g_5^2\pi R}$. We can also define the canonically normalized axion field by $a_5(x)\equiv f_a\theta_5(x)$
where the subscript for $\theta_5, a_5$ is not the spatial index but denotes that they are defined in the 5D picture.
Keeping only the zero modes which are independent of the $y$-coordinate in the CS term and integrating over the fifth dimension give us the familiar form of the axion-gluon coupling,
\beq \mathcal{L}_{agg}=-\frac{\kappa_{\rm CS}}{16\pi^2}\frac{a_5(x)}{f_a} \tr \lmk {\mathcal{G}}_{\mu\nu,0} (x)\widetilde{\mathcal{G}}^{\mu\nu}_0 (x) \rmk\, ,\eeq
where $\mathcal{G}_{\mu\nu,0}, \widetilde{\mathcal{G}}^{\mu\nu}_0$ are the field strength of the massless gluon
and its dual. 

As we have mentioned, the 5D gauge couplings $g_5,g_{5,c}$ have negative mass dimensions and the theory is non-renormalizable.
Therefore, we have a UV cutoff scale where the gauge coupling becomes strong,
\begin{align}
    \Lambda_5 \lesssim \mathrm{min}\left\{\frac{24\pi^3}{g_5^2},~\frac{24\pi^3}{g_{5,c}^2}\right\}\, ,
\end{align}
where $24\pi^3$ is the loop factor in 5D.
Motivated by this, we will consider a renormalizable four-dimensional field theory that can reproduce the properties of the five-dimensional gauge theory and potentially become its UV completion.

For the convenience of later discussions on the axion potential, we introduce a bulk fermion charged under both $U(1)$ and $SU(3)$ as well as a bulk scalar field charged under only $U(1)$. Let us first consider the bulk fermion $\Psi(x,y)$ which has the $U(1)$ charge $+1$ and transforms as the fundamental representation under $SU(3)$. The action can be written as
\begin{align}
    S_{5}^{f}=\int  \dd^5 x \lkk\overline{\Psi}(x,y)i\Gamma^M D_M \Psi(x,y) - M_{\Psi}\overline{\Psi}(x,y)\Psi(x,y)\rkk\, .
\end{align}
Here, $M_\Psi$ is the bulk mass, $\overline{\Psi}=\Psi^\dag \Gamma^0$, $D_M \Psi= \del_M\Psi + iA_M\Psi + iG_M\Psi$
is the covariant derivative and the 5D Dirac matrices include 
\begin{align}
	\Gamma^\mu=\gamma^\mu=\begin{pmatrix}
	0&\sigma^\mu\\ \overline{\sigma}^\mu &0
\end{pmatrix}, \quad \Gamma^5= -i\gamma^5= \gamma^0\gamma^1\gamma^2\gamma^3=i\begin{pmatrix}
	I&0\\0&-I
\end{pmatrix}\, .
\end{align}
Here, $\sigma^\mu = (1,\vec{\sigma})$ and $\bar{\sigma}^\mu = (1,-\vec{\sigma})$ with the Pauli matrices $\vec{\sigma}$.
We can expand $\Psi$ in terms of two Weyl components as
\begin{align}
    \Psi=\lmk \eta_\alpha,\psi^{\dag\dot\alpha}\rmk^T\, ,
\end{align}
where $\eta~(\psi)$ transforms as the (anti-)fundamental representation under $SU(3)$ and carries $U(1)$ charge $+1~(-1)$. 
Then, the action can be expressed as
\begin{align} S_{5}^f =& \int \dd^4x \int_0^{\pi R}\dd y \lkk\eta^\dag(x,y) i\overline{\sigma}^\mu D_\mu \eta(x,y) + \psi^\dag(x,y) i\overline{\sigma}^\mu D^*_\mu \psi(x,y)\right.\notag \\
&\left. -\psi(x,y) D_5\eta(x,y) +\eta(x,y)^\dag D_5 \psi(x,y)^\dag -M_\Psi (\psi\eta+\mathrm{h.c.})\rkk \label{5D_fermion}\, .\end{align}
We impose the Dirichlet boundary conditions on $\psi(x,y)$,\footnote{The boundary conditions on $\eta(x,y)$ are determined by the bulk equations mixing two Weyl fermions. The detailed explanation is given in Appendix~\ref{sec:appendix_5D_spectrum}.}
\begin{align}
   \delta\psi|_{0,\pi R} = 0\, , \qquad \psi|_{0,\pi R}=0\, , \label{fermion_BC}
\end{align}
where the second equality follows directly from the 4D Lorentz invariance on the boundaries.
Note that the above boundary conditions ensure that the kinetic terms in Eq.~\eqref{5D_fermion} are Hermitian.
We perform the KK expansion of the Weyl fermions,
\begin{align}
    \eta(x,y)=\sum_{n=0}^{\infty}\eta_n(x)f_n^\eta(y) \, , \qquad \psi(x,y)=\sum_{n=0}^{\infty}\psi_n(x)f_n^\psi(y)\, , \label{KK_expansion_fermion}
\end{align}
and the profiles can be solved as
\begin{align}
    f_0^\psi = 0 \, &,\quad f_n^\psi = \sqrt{\frac{2}{\pi R}} \sin\frac{ny}{R} \quad \text{for} \,\,\, n\geq 1\, , \\[1ex]
    f_0^{\eta} = N_0 e^{-M_\Psi y}\, &, \quad f_n^\eta = \sqrt{\frac{2}{\pi R}} \lmk \frac{M_\Psi}{m_n}\sin\frac{ny}{R} - \frac{n}{m_n R}\cos\frac{ny}{R} \rmk \quad \text{for} \,\,\, n\geq 1\, , 
\end{align}
with $m_{f,n}^2 = M_\Psi^2 +\frac{n^2}{R^2}$.
The normalization factor of the zero mode is given by $N_0=\left(\int_0^{\pi R}\dd y\,e^{-2M_\Psi y}\right)^{-1/2}$.
Note that $\Psi$ has a chiral zero mode which comes from $\eta$. The KK spectrum is explicitly given by
\begin{align}
&\text{Zero mode:}\qquad \eta_0(x) = N_0\int_0^{\pi R}\dd y\, e^{-M_\Psi y}\eta(x,y)\, ,\\[1ex]
&\text{Dirac fermion pairs with mass squared $m_{f,n}^2=M_\Psi^2+\frac{n^2}{R^2}$:} \nonumber \\
&~~\eta_n(x) = \int_0^{\pi R}\dd y\, f_n^\eta(y)\eta(x,y)\, ,\quad 
\psi_n(x) = \int_0^{\pi R}\dd y\, f_n^\psi(y)\psi(x,y)\quad \text{for} \,\,\, n\geq 1\, . \label{fermion_KK_profile}
\end{align}
It is known that integrating out the KK modes of such bulk fermions will lead to the 5D CS term~\eqref{5D_CS}~\cite{Bonnefoy:2020llz, Arkani-Hamed:2001uol}, and we will see the corresponding discussion in the deconstruction picture in Sec.~\ref{sec:deconstruction}.

Next, we consider a bulk complex scalar $Q(x,y)$ which has the $U(1)$ charge $+1$. The action is written as
\begin{align}
	S_5^s = \int\dd^4 x &\int_0^{\pi R}\dd y\lmk \lkk D_M Q(x,y)\rkk^* D_M Q(x,y) - M_Q^2 \abs{Q(x,y)}^2 \rmk \, ,\label{5D_scalar}
\end{align}
with $D_M Q = \del_M Q +i A_M Q$.
We assume the Neumann boundary conditions for $Q(x,y)$,
\begin{align}
	\del_y Q(x,y)|_{0,\pi R} = 0\, .\label{scalar_BC}
\end{align}
Then, the KK spectrum of $Q(x,y)$ is given by
\begin{align}
	&f_0^Q(y) =\sqrt{\frac{1}{\pi R}}\, ,\quad  f_n^Q(y) = \sqrt{\frac{2}{\pi R}} \cos\frac{ny}{R}\, ,\quad n>0\, , \label{scalar_KK_spectrum} \\[1ex]
    &Q_n(x) = \int_0^{\pi R} \dd y\,f_n^Q(y)Q(x,y)\, ,\quad n\geq 0\, ,
\end{align}
where the KK modes are defined as $Q(x,y)=\sum_{n=0}^\infty Q_n(x)f_n^Q(y)$ and
have mass squared $m_{Q,n}^2 = M_Q^2 + \frac{n^2}{R^2}$.
The derivation of the above spectrum can be found in Appendix~\ref{sec:appendix_5D_spectrum}.

\section{Deconstruction
\label{sec:deconstruction}}
The deconstruction setup provides a four-dimensional realization of a five-dimensional gauge theory, where the fifth dimension emerges as a discrete lattice.
The continuous coordinate $0\leq y\leq \pi R$ is discretized to $y_j\equiv ja, ~1\leq j\leq N$ with $Na=\pi R$ and the integration $\int_0^{\pi R}\dd y$ is replaced by a collection of lattice sites: the summation $\sum_{i=1}^N a$.

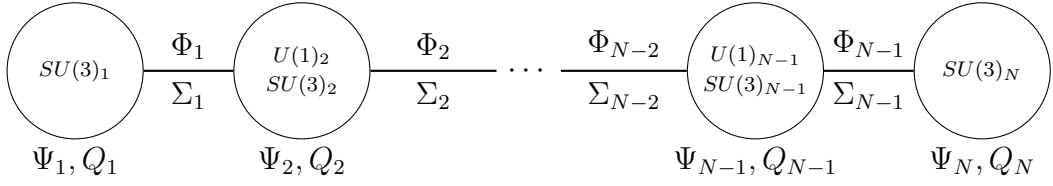
\begin{figure}[!t]
\centering
\begin{tikzpicture}[
    site/.style={
        circle,
        draw,
        minimum width=1.8cm,
        minimum height=1.8cm,
        inner sep=2pt,
        align=center,
        font=\scriptsize
    }
]
\node[site] (G1) at (0,0) {$SU(3)_1$};

\node[site] (G2) at (3,0)
{\shortstack{$U(1)_2$\\ $SU(3)_2$}};

\node (G3) at (6,0) {$\cdots$};

\node[site] (Gm) at (9,0)
{\shortstack{$U(1)_{N-1}$\\ $SU(3)_{N-1}$}};

\node[site] (GN) at (12,0) {$SU(3)_N$};

\draw[thick] (G1) -- node[above] {$\Phi_1$}
                     node[below] {$\Sigma_1$} (G2);

\draw[thick] (G2) -- node[above] {$\Phi_2$}
                     node[below] {$\Sigma_2$} (G3);

\draw[thick] (G3) -- node[above] {$\Phi_{N-2}$}
                     node[below] {$\Sigma_{N-2}$} (Gm);

\draw[thick] (Gm) -- node[above] {$\Phi_{N-1}$}
                     node[below] {$\Sigma_{N-1}$} (GN);

\node at (0,-1.2) {$\Psi_1,Q_1$};
\node at (3,-1.2) {$\Psi_2,Q_2$};
\node at (9,-1.2) {$\Psi_{N-1},Q_{N-1}$};
\node at (12,-1.2) {$\Psi_N,Q_N$};

\end{tikzpicture}
\caption{Schematic illustration of the deconstructed $U(1)\times SU(3)$ setup. 
Adjacent sites are connected by the link fields $\Phi_j$ and $\Sigma_j$, while the charged matter fields $\Psi_j$ and $Q_j$ are localized at the sites.
The absence of the $U(1)$ gauge groups at the endpoints corresponds to the Dirichlet boundary condition for $A_\mu(x,y)$ in 5D.}
\label{setup_fig}
\end{figure}

\subsection{Gauge Sector}
Let us first present the deconstructed action of the 5D $U(1)$ gauge field with the Dirichlet/Neumann boundary conditions on $A_\mu/A_5$ components,
\beq S^{U(1)}_D = \int\dd^4x\lmk - \frac{1}{4g^2} \sum_{j=2}^{N-1} \mathcal{F}_j^{\mu\nu}\mathcal{F}_{j,\mu\nu}+\sum_{j=1}^{N-1}D_\mu\Phi_j^* D^\mu\Phi_j-V(\Phi)\rmk\, ,\label{deconst_U1}\eeq
where we accommodate one $U(1)_j$ gauge field $A_j^\mu$ on each site, $j=2,3,\cdots,N-1$, with the same coupling constant $g$.
The absence of gauge fields on the endpoints $j=1,N$ corresponds to the Dirichlet boundary condition for $A_\mu(x,y)$ in 5D. 
Between the $j$-th and $(j+1)$-th sites, we have one complex scalar field $\Phi_j$ which has charge $+1$ and $-1$ under $U(1)_j$ and $U(1)_{j+1}$, called a link field ($\Phi_1$ and $\Phi_{N-1}$ are charged only under $U(1)_2$ and $U(1)_{N-1}$ respectively).  
Fig.~\ref{setup_fig} gives a schematic picture of the deconstruction setup and Tab.~\ref{tab:charge} shows the charge assignment for the link fields.

Each link scalar field acquires a common VEV $\vev{\Phi_j}=\frac{v}{\sqrt{2}}$ via the potential $V(\Phi)$ in Eq.~\eqref{deconst_U1} and all the $U(1)_j$ gauge symmetries are spontaneously broken.
Corresponding to the translational invariance along the fifth dimension, the gauge couplings and the VEVs of the link fields are identical across the lattice.
To correctly reproduce the 5D theory in the continuum limit, the interval $a$ is interpreted as the product of the coupling constant and the symmetry breaking scale $gv$.
The complete correspondence between the deconstruction and 5D pictures is derived as follows: 
\begin{align} 
a\equiv \frac{\pi R}{N} &\longleftrightarrow (gv)^{-1}\, ,\notag\\
 y_j/a &\longleftrightarrow j\, ,\notag\\
 A_\mu(x,y_j) &\longleftrightarrow A_{j,\mu}(x)\, ,\notag\\
 A_5(x,y_j) &\longleftrightarrow g\chi_j(x)\, ,\notag\\
 \frac{a}{g_5^2} &\longleftrightarrow \frac{1}{g^2} \, ,\label{correspondence}
\end{align}
where the continuum limit is achieved by taking $N\rightarrow\infty$ with the length $\pi R$ and the 5D coupling $g_5$ fixed, equivalently, $v\rightarrow\infty$. 
For simplicity, we neglect the radial modes of the link fields and identify $\Phi_j$ as $\frac{v}{\sqrt{2}}e^{i\chi_j/v}$. 
The action is reduced to:
\begin{align}
S_D^{U(1)}&=\int\dd^4 x \lmk  -\frac{1}{4g^2} \sum_{j=2}^{N-1}\mathcal{F}_j^{\mu\nu}\mathcal{F}_{j,\mu\nu} +\frac{1}{2}v^2\lkk \lmk \del_\mu\chi_1/v-A_{2,\mu}\rmk^2+\lmk\del_\mu\chi_{N-1}/v+A_{N-1,\mu}\rmk^2\rkk\right.\notag\\
&\left. +\frac{1}{2}v^2\sum_{j=2}^{N-2}\lmk\del_\mu\chi_j/v +A_{j,\mu}-A_{j+1,\mu}\rmk^2\rmk\notag \\
& = \int\dd^4 x \lmk -\frac{1}{4g^2} \sum_{j=2}^{N-1}\mathcal{F}_j^{\mu\nu}\mathcal{F}_{j,\mu\nu}+\frac{1}{2g^2}\sum_{j=1}^{N-1}\lkk g\del_\mu \chi_j + gv(A_{j,\mu}-A_{j+1,\mu})\rkk^2\rmk \label{deconst_U1_2}\\
	&\leftrightarrow \int\dd^4 x \lmk -\frac{a}{4g_5^2} \sum_{j=2}^{N-1}\mathcal{F}^{\mu\nu}\mathcal{F}_{\mu\nu}(y_j)+\frac{a}{2g_5^2}\sum_{j=1}^{N-1}\lkk \del_\mu A_5(y_j) - \frac{A_{\mu}(y_j+a)-A_{\mu}(y_j)}{a}\rkk^2\rmk\, ,
\end{align}
where we can see that in the continuum limit, the 5D action is correctly reproduced after applying the correspondence~\eqref{correspondence}. In the last term of Eq.~\eqref{deconst_U1_2}, we formally introduced $A_{1,\mu}, A_{N,\mu}$ which are always set to zero.

\begin{table}[t!]
  \begin{tabular}{c|cccc}
            &$U(1)_j$ & $U(1)_{j+1}$ & $SU(3)_j$ & $SU(3)_{j+1}$\\
    \hline
    $\Phi_j$& $+1$    & $-1$&$\1$ & $\1$\\
    $\Sigma_j$   & $0$     & $0$& $\square$ &$\overline{\square}$\\
    $\Psi_j$& $+1$& $0$&$\square$ &$\1$\\
    $Q_j$ & $+1$ & $0$ & $\1$ & $\1$ 
  \end{tabular}
  \caption{Charge assignments in the deconstruction setup.  For the link fields $\Phi_j, \Sigma_j$, $j = 1,\dots, N-1$; for the matter fields $\Psi_j,Q_j$, $j = 1,\dots, N$. 
  Due to Dirichlet boundary conditions, only $2 \le j \le N-1$ have $U(1)_j$. 
  For the boundary fields, $\Phi_1$ and $\Phi_{N-1}$ carry charges $-1$ and $+1$ under $U(1)_2$ and $U(1)_{N-1}$, respectively, while $\Psi_1,Q_1$ and $\Psi_N,Q_N$ are neutral under all $U(1)$.}
  \label{tab:charge}
\end{table}

We can directly read the gauge field mass term from Eq.~\eqref{deconst_U1_2},
\begin{align}
	\mathcal{L}_{gm}&=\frac{1}{2} v^2\lkk A_{2,\mu}^2+\sum_{j=2}^{N-2}\lmk A_{j+1,\mu}-A_{j,\mu}\rmk^2+A_{N-1,\mu}^2\rkk  \notag \\
   & =  \frac{1}{2} g^2 v^2\lkk \hat{A}_{2,\mu}^2+\sum_{j=2}^{N-2}\lmk \hat{A}_{j+1,\mu}-\hat{A}_{j,\mu}\rmk^2+\hat{A}_{N-1,\mu}^2\rkk  \nonumber \\
   & \equiv \frac{1}{2} \sum_{i,j} \hat{A}_{i,\mu} \lmk \mathcal{M}_{U(1)}^2\rmk_{ij}\hat A_{j,\mu} \, .
   \label{A_mass}
\end{align}
Here, we have introduced $\hat{A}_{j,\mu}=A_{j,\mu}/g$ with canonically normalized kinetic terms to solve for the eigensystem of the mass matrix $\mathcal{M}_{U(1)}^2$.
Let us define the discrete Laplacian matrices with Dirichlet/Neumann boundary conditions on the two ends as
\begin{align}
    \Delta^{\rm D}_N = \begin{pmatrix}
2 & -1 & 0 & \cdots & 0\\
-1 & 2 & -1 & \cdots & 0\\
0 & -1 & 2 & \ddots & \vdots\\
\vdots & & \ddots & \ddots & -1\\
0 & 0 & \cdots & -1 & 2
\end{pmatrix} \, ,\qquad
\Delta^{\rm N}_N =
\begin{pmatrix}
1 & -1 & 0 & \cdots & 0\\
-1 & 2 & -1 & \cdots & 0\\
0 & -1 & 2 & \ddots & \vdots\\
\vdots & & \ddots & \ddots & -1\\
0 & 0 & \cdots & -1 & 1
\end{pmatrix} \, , \label{Delta_N}
\end{align}
where the superscripts $\mathrm{D,N}$ denote the type of boundary conditions while the subscript $N$ denotes the dimension of the square matrix.
Then, the mass matrix can be written as
\begin{align}
    \mathcal{M}_{U(1)}^2 = g^2 v^2 \Delta_{N-2}^{\rm D}\, .
\end{align}
The mass eigenvalues and eigenstates are solved as
\begin{align}
	&m_k^2 = 4g^2v^2\sin^2\lmk \frac{k\pi}{2N-2}\label{eigenmass_U1}\rmk\, ,\\[1ex]
	&\hat{A}_\mu^{(k)}=\sqrt{\frac{2}{N-1}}\sum_{j=2}^{N-1}\sin\lmk \frac{(j-1)k\pi}{N-1}\rmk \hat{A}_{j,\mu}\, ,~~\text{with}~~k=1,\cdots,N-2\, , \label{eigenstate_U1_canonical}
\end{align}
where the normalization in Eq.~\eqref{eigenstate_U1_canonical} is set by the canonical normalization of the $\hat{A}_\mu^{(k)}$ kinetic terms.
Then we switch back to the convention in which the gauge fields absorb the gauge coupling. 
It is somewhat subtle that the mass eigenstates should absorb the low-energy effective gauge coupling, which is defined from the correspondence with the 5D picture as
\begin{align}
    \frac{1}{g_D^2}=\frac{N-1}{g^2} \longleftrightarrow \frac{(N-1)a}{g_5^2} \simeq \frac{\pi R}{g_5^2} = \frac{1}{g_4^2}\, .
\end{align}
Therefore we have the relation $A_\mu^{(k)} = g_D\hat{A}_{\mu}^{(k)} = \frac{1}{\sqrt{N-1}}g \hat{A}_{\mu}^{(k)}$ and finally,
\begin{align}
    A_\mu^{(k)} =\frac{\sqrt{2}}{N-1}\sum_{j=2}^{N-1}\sin\lmk \frac{(j-1)k\pi}{N-1}\rmk A_{j,\mu}\, .
\end{align}
Following the similar linear-algebraic procedure, we can determine the linear combinations of the  Goldstone bosons $\chi_j$ that are eaten by the gauge fields through the Higgs mechanism,
\begin{align}
	\chi^{(k)}=\sqrt{\frac{2}{N-1}}\sum_{j=1}^{N-1}\cos\lmk \frac{(j-\half)k\pi}{N-1}\rmk \chi_{j}\, ,~~\text{with}~~k=1,\cdots,N-2\, ,
\end{align}
where $\chi^{(k)}$ is eaten by $A_\mu^{(k)}$ correspondingly to form a massive gauge boson.
One linear combination is left as a massless Goldstone boson in the theory, which is identified as the axion,
\begin{align} a(x)\equiv \frac{1}{\sqrt{N-1}}\sum_{j=1}^{N-1}\chi_j(x)\, .\label{deconst_axion}\end{align}
This can be checked by looking at the mixing term in Eq.~\eqref{deconst_U1_2},
\begin{align}
    &\frac{1}{2}\sum_{j=1}^{N-1} \lmk \del_\mu\chi_j + gv \hat A_{j,\mu}-gv \hat A_{j+1,\mu}\rmk^2 \notag \\
    = &\frac{1}{2}\sum_{j=1}^{N-1} \lmk \del_\mu\chi_j + gv \sqrt{\frac{2}{N-1}}\sum_{k=1}^{N-2}\lkk \sin\lmk\frac{(j-1)k\pi}{N-1}\rmk- \sin\lmk\frac{jk\pi}{N-1}\rmk\rkk \hat A^{(k)}_{\mu}\rmk^2 \notag \\
    =& \frac{1}{2}\sum_{j=1}^{N-1} \lmk \del_\mu\chi_j -\sqrt{\frac{2}{N-1}}\sum_{k=1}^{N-2}  m_k \cos\lmk\frac{(j-\half) k\pi}{N-1}\rmk \hat A^{(k)}_{\mu}\rmk^2 \notag \\
    =& \frac{1}{2} \lmk \del_\mu a \rmk^2 + \half \sum_{k=1}^{N-2}\lmk \del_\mu \chi^{(k)} - m_k \hat A_\mu^{(k)}\rmk^2 \, . \label{U1_Higgs_mechanism}
\end{align}
In the last equality, we have utilized the inverse relation between $\chi^{(k)}$ and $\chi_j$ and the orthogonality of cosine functions.
According to the correspondence~\eqref{correspondence}, Eqs.~\eqref{eigenmass_U1}-\eqref{deconst_axion} reproduce the complete KK spectrum of $A_M$~\eqref{KKspec_A50},~\eqref{KKspec_AG} in the continuum limit.

Similarly, the deconstructed action of the $SU(3)$ gauge theory is given by
\beq S^{SU(3)}_D = \int\dd^4x\lmk -\frac{1}{2g_c^2} \sum_{j=1}^{N}
\tr\lmk\mathcal{G}_j^{\mu\nu}\mathcal{G}_{j,\mu\nu}\rmk+\sum_{j=1}^{N-1}\tr \lmk\lmk D_\mu \Sigma_j\rmk^\dag D^\mu \Sigma_j\rmk-V(\Sigma)\rmk\, ,\label{deconst_SU3}\eeq
where the link fields $\Sigma_j$ follow the charge assignment in Tab.~\ref{tab:charge} and thus $D_\mu\Sigma_j=\del_\mu\Sigma_j+i G_{j,\mu}\Sigma_j-i\Sigma_j G_{j+1,\mu}$. Because of the Neumann boundary conditions on the $G_\mu$ component, at the endpoints $G_1(x),G_N(x)$ are included. The potential $V(\Sigma)$ is set in the way that the link fields acquire a common VEV $\vev{\Sigma_j}=v_3$ and spontaneously break all the $SU(3)$ gauge groups except for the diagonal subgroup $SU(3)^{(0)}$. With the simplified link fields $\Sigma_j\simeq v_3 \exp\lmk i\pi_j/v_3 \rmk\equiv v_3 \exp\lmk i\pi_j^a T^a/v_3\rmk$, we have the correspondence,
\begin{align}
	G_\mu(x,y_j)\longleftrightarrow G_j(x) \, , \quad G_5(x,y_j)\longleftrightarrow g_c\pi_j \, , \quad g_c v_3\longleftrightarrow a^{-1}\, .
\end{align}
We require $g_c v_3= g v$ for the consistency with deconstruction of the $U(1)$ gauge theory.
Substituting $\Sigma_j\simeq v_3 \exp\lmk i\pi_j/v_3\rmk$ into the action~\eqref{deconst_SU3}, we obtain the mass matrix of the canonically normalized fields $\hat G_{j,\mu}\equiv G_{j,\mu}/g_c$,
\begin{align}
    \mathcal{M}_{SU(3)}^2 = g_c^2 v_3^2 \Delta_{N}^{\rm N}\, .
\end{align}
The mass spectrum is solved to be
\begin{align}
	&m_k^2 = 4g_c^2v_3^2\sin^2\lmk \frac{k\pi}{2N}\rmk\, ,\\[1ex]
	&G_\mu^{(k)}=\frac{\sqrt{c_k}}{N}\sum_{j=1}^{N}\cos\lmk \frac{(j-\half)k\pi}{N}\rmk G_{j,\mu}\, ,~~\text{with}~~ k=0,1,\cdots,N-1\, , \label{eigenstate_su3}
\end{align}
where $c_k=1$ for $k=0$ and $c_k=2$ for $k\geq1$. Here the results are directly shown in the convention where the gauge fields absorb the low-energy effective $SU(3)$ gauge coupling, defined as
\begin{align}
	\frac{1}{g_{c,D}^2}=\frac{N}{g_c^2}\longleftrightarrow \frac{Na}{g_{5,c}^2}=\frac{\pi R}{g_{5,c}^2}=\frac{1}{g_{4,c}^2}\, .
\end{align}
This also reproduces the KK spectrum~\eqref{KKspec_Gmu0}, \eqref{KKspec_AG} when $N\rightarrow\infty$.
The massless zero mode $G^{(0)}_\mu=\frac{1}{N}\sum_{j}G_{j,\mu}$ is the gauge field of the diagonal subgroup $SU(3)^{(0)}$ remaining at low energies.
The effective gauge coupling $g_{c,D}$ just coincides with the coupling constant of $SU(3)^{(0)}$,
\begin{align}
    \frac{1}{g^2_{c,D}} = \frac{1}{g_{c,\rm diag}^2} = \frac{N}{g_c^2}\, , 
\end{align}
where the last equality originates from the symmetry breaking pattern $SU(3)_1\times SU(3)_2\times\cdots\times SU(3)_N\longrightarrow SU(3)^{(0)}$.

As mentioned in Sec.~\ref{sec:review}, a five-dimensional gauge theory is non-renormalizable in general and has a UV cutoff $\Lambda_5$, defined as
\begin{align}
    \Lambda_5\lesssim \frac{24\pi^3}{g_5^2}\, .
\end{align}
Here, we use $g_5,\Lambda_5$ and $g,v$ to represent the quantities of a general 5D gauge theory and their deconstruction counterparts, rather than referring specifically to the $U(1)$ sector.
The deconstruction picture, as a four-dimensional field theory, can potentially provide a UV completion of the 5D gauge theory.
Then the limit of $v\to\infty$ should be regarded as a formal idealization since the physical VEV of the link fields certainly cannot go to infinity.
Therefore, the continuum limit should be considered more carefully.
Notice that the mass spectrum in the deconstruction picture, like Eq.~\eqref{eigenmass_U1}, clearly deviates from that in the 5D picture when $k$ becomes comparable to $N$.
This gives a requirement for the deconstruction setup to be a successful UV completion: all the deconstruction eigenmasses below the 5D cutoff $\Lambda_5$ should be approximated to the KK masses.
Define the critical level of the deconstruction eigenmass $k_c$ as
\begin{align}
  m_{k_c}<\Lambda_5\, ,\quad m_{k_c+1}>\Lambda_5\, , 
\end{align}
which means $m_{k_c}$ is the largest eigenmass below the 5D cutoff scale.
The requirement follows
\begin{align}
    m_{k_c} = 2g v \sin\lmk\frac{k_c\pi}{2N}\rmk \simeq \frac{k_c}{R} \, ,
\end{align}
which is equivalent with $k_c\ll N/\pi$.
Approximately taking $\Lambda_5\simeq m_{k_c}$, we have
\begin{align}
    \Lambda_5\simeq \frac{k_c}{R} \ll \frac{N}{\pi R} = \frac{1}{a} =gv\, . \label{UV_completion_condition}
\end{align}

\subsection{Matter Sector
\label{sec:Fermion_Deconst}}
The deconstructed action of the bulk fermion $\Psi$ can be written in terms of the Weyl components $\psi,\eta$ as
\begin{align}
	S_{D}^f =& \int\dd^4 x\lkk \sum_{j=1}^N  \eta_j^\dag i\overline{\sigma}^\mu D_\mu\eta_j+\sum_{j=1}^{N-1}\lmk \psi_j^\dag i\overline{\sigma}^\mu D^*_\mu\psi_j-M_\Psi\lmk\eta_j\psi_j+\psi_j^\dag\eta_j^\dag\rmk\rmk\right. \notag\\
  &\left. -\sum_{j=1}^{N-1}\lmk \frac{\sqrt{2}g}{v_3}\psi_j \Phi_j\Sigma_j\eta_{j+1} 
  -gv\eta_j\psi_j +\mathrm{h.c.}\rmk\rkk \label{deconst_fermion_0}\\
  =& \int\dd^4 x\lkk  \mathcal{L}_{\rm kin}-\sum_{j=1}^{N-1} \lmk M'_\Psi\eta_j\psi_j + \frac{\sqrt{2}g}{v_3}\psi_j \Phi_j\Sigma_j\eta_{j+1} +\mathrm{h.c.}\rmk \rkk\, , \label{deconst_fermion}
\end{align}
with the effective Dirac mass $M_\Psi'=M_\Psi-gv$~\cite{Skiba:2002nx,Abe:2002rj}.
In this action, we put fermions $\eta_j,\psi_j$ on each lattice site but remove $\psi_N$ to reproduce the Dirichlet boundary condition~\eqref{fermion_BC}. The charge assignment is shown in Tab.~\ref{tab:charge}. Since in the 4D theory, the mass dimension of the spinor fields is $\lkk \eta_j(x)\rkk=\lkk \psi_j(x)\rkk=\frac{3}{2}$, we have the correspondence,
\begin{align}
	a^{1/2}\psi(x,y_j) \longleftrightarrow \psi_j(x)	~&, \quad a^{1/2}\eta(x,y_j) \longleftrightarrow \eta_j(x)\, .
\end{align}
Then it is easy to see that the first line in Eq.~\eqref{deconst_fermion_0} can reproduce the 4D covariant derivatives and the bulk mass term in Eq.~\eqref{5D_fermion} in the continuum limit. Let us check the second line,
\begin{align}
	-\sum_{j=1}^{N-1}&\lmk \frac{\sqrt{2}g}{v_3}\psi_j \Phi_j\Sigma_j\eta_{j+1} -gv\eta_j\psi_j +\mathrm{h.c.}\rmk 
  =-(gv)^2\sum_{j=1}^{N-1}\lmk\psi_j \frac{\tilde{\Phi}_j\tilde{\Sigma}_j\eta_{j+1} 
  -\eta_j}{gv} +\mathrm{h.c.}\rmk \notag\\
  &\qquad \longleftrightarrow - \sum_{j=1}^{N-1} a \lmk \psi(x,y_j) D_5^+ \eta(x,y_j)+ \mathrm{h.c.} \rmk\, ,
\end{align}
where 
\begin{align}
	\tilde{\Phi}_j\equiv \frac{\sqrt{2}\Phi_j}{v}=e^{i\chi_j/v}\leftrightarrow e^{iaA_5}
	, \quad \tilde{\Sigma}_j\equiv \frac{\Sigma_j}{v_3}=e^{i\pi_j/v_3}\leftrightarrow e^{iaG_5}\, ,
\end{align} 
correspond to the link variables defined in the lattice gauge theory and give the forward covariant derivative in the fifth direction, 
\begin{align}D_5^+\eta(x,y_j)=\frac{e^{iaA_5}e^{iaG_5}\eta(x,y_{j+1})-\eta(x,y_j)}{a}\, .\end{align}
Therefore, the 5D action of the bulk fermion is correctly reproduced in the continuum limit.

One may notice that the interaction of $\psi_j,\eta_{j+1}$ with the link fields, usually called as the hopping term, has mass dimension larger than $4$, which is a general feature for the matter fields charged under multiple gauge groups.
Since we are trying to construct a renormalizable theory, we give one possible UV completion of such non-renormalizable term here.
We introduce one heavy scalar mediator $S_j~(j=1,\cdots,N-1)$  with mass $M_S$ for each pair of the link fields, which has the same charge as the product of the two link fields.
Therefore, it can couple with the fermions and the link fields by the renormalizable interactions,
\begin{align}
    \mathcal{L}_{S}\supset -y_S \psi_j S_j\eta_{j+1}  - \mu_S S_j^\dag\Sigma_j\Phi_j +\mathrm{h.c.} \, ,
\end{align}
where $y_S$ is dimensionless and $\mu_S$ has mass dimension $1$.
Integrating out $S_j$, we obtain an effective interaction at the tree level,
\begin{align}
    \mathcal{L}_{S,\rm eff} \supset -\frac{y_S\mu_S}{M_S^2} \psi_j \Phi_j\Sigma_j\eta_{j+1} + \mathrm{h.c.}
\end{align}
To match the hopping term in the deconstructed action, we require the parameters to have the relation,
\begin{align}
    \frac{y_S\mu_S}{M_S^2} \simeq \frac{\sqrt{2}g}{v_3} \, .
\end{align}

Now let us check the mass spectrum. The fermion mass terms are given by
\begin{align}
	\mathcal{L}_{fm} = -\sum_{j=1}^{N-1}\lmk M_\Psi'\psi_j\eta_j + gv \psi_j\eta_{j+1}+\mathrm{h.c.}\rmk\equiv -\sum_{j=1}^{N-1}\sum_{k=1}^N \psi_j \lmk\mathcal{M}_f\rmk_{jk}\eta_k\, +\mathrm{h.c.} ,
\end{align}
where the $(N-1)\times N$ mass matrix is written as
\begin{align}
	\mathcal{M}_f=\begin{pmatrix}
		M_\Psi' & gv & 0 & \cdots & 0 &0\\
		0 & M_\Psi' & gv & \cdots & 0 &0\\
		0 & 0 & M_\Psi' & \ddots & 0 &0\\
		\vdots & \vdots & \vdots &\ddots & \ddots& \vdots\\
		0 &0 & 0 &\cdots & M_\Psi' & gv
	\end{pmatrix}\, .
\end{align}
The mass eigenstates are given by the eigenvectors of the Hermitian mass matrix squared $\mathcal{M}_f\mathcal{M}_f^\dag$ acting on $\psi$ and $\mathcal{M}_f^\dag \mathcal{M}_f$ acting on $\eta$.
The result is shown as
\begin{align}
	&\text{Zero mode:}~~ \eta^{(0)}=\mathcal N_0 \sum_{j=1}^{N} \left(-\frac{M'_\Psi}{gv}\right)^{j-1}\eta_j \, ,
\qquad |\mathcal N_0|^{-2} = \sum_{j=1}^{N}\left|\frac{M_\Psi'}{gv}\right|^{2(j-1)}\, , \\[1ex]
&\text{Massive Dirac fermions with mass squared $m_n^2 = M_\Psi'^2+(gv)^2+2M_\Psi'gv\cos\!\left(\frac{n\pi}{N}\right)$:} \nonumber \\[1ex]
&\eta^{(n)}=\sqrt{\frac{2}{N}}\sum_{j=1}^{N}\frac{1}{m_n}\left[ M'_\Psi\sin\!\left(\frac{j n\pi}{N}\right)+gv\sin\!\left(\frac{(j-1)n\pi}{N}\right)\right]\eta_j \, ,\\[1ex]
&\psi^{(n)} =\sqrt{\frac{2}{N}}\sum_{j=1}^{N-1}\sin\left(\frac{j n\pi}{N}\right)\psi_j\, , \qquad n=1,\dots,N-1 \, .
\end{align}
The mass eigenvalues reproduce the KK mass tower in the continuum limit since
\begin{align}
	m_n^2 =M_\Psi^2+4[(gv)^2 -M_\Psi(gv)]\sin^2\lmk\frac{n\pi}{2N}\rmk \leftrightarrow M_\Psi^2+\frac{n^2}{R^2} \, ,
\end{align}
where the term linear in $gv$ vanishes in the limit of $N\rightarrow \infty$.
The mass eigenstates of $\psi_j$ match the KK wavefunction in Eq.~\eqref{fermion_KK_profile}.
Let us demonstrate that the mass eigenstates of $\eta_j$ also reproduce the KK profiles in the continuum limit.
For the zero mode,
\begin{align}
    \eta^{(0)} \propto \sum_{j=1}^N \lmk 1 - \frac{M_\Psi}{gv}\rmk^{j-1}\eta_j \simeq \sum_{j=1}^N \exp \lmk - (j-1)\frac{M_\Psi}{gv}\rmk\eta_j \longleftrightarrow \sum_{j=1}^N \exp \lmk - M_\Psi y_j\rmk\eta_j \, ,\notag
\end{align}
where we use the relation $(1-x)^n\simeq 1-nx\simeq e^{-nx}$ for small $x$ and neglect the mismatch of one unit in the last correspondence.
The normalization factor $\mathcal{N}_0$ can be matched in a similar way.
For the massive modes, expanding $M_\Psi'=M_\Psi-gv$, we have a term inside the bracket as
\begin{align}
    &gv\lkk \sin\lmk\frac{(j-1)k\pi}{N}\rmk- \sin\lmk\frac{jk\pi}{N}\rmk\rkk = -2gv\sin\frac{n\pi}{2N}\cos\lmk\frac{(j-\half)n\pi}{N} \rmk\notag \\
    &\simeq -2 gv\frac{n\pi}{2N}\cos\lmk\frac{(j-\half)n\pi}{N}\rmk \longleftrightarrow -\frac{n}{R}\cos\lmk\frac{(j-\half)n\pi}{N}\rmk \, .
\end{align}
After this identification, the matching with Eq.~\eqref{fermion_KK_profile} becomes clear.
 
For the complex scalar $Q$, first neglecting the brane-localized operators, the deconstructed action is given by 
\begin{align}
{S_D^s} &=\int\dd^4 x\lkk  \sum_{j=1}^N \lmk (D_\mu Q_j)^* D^\mu Q_j-M_Q^2 |Q_j|^2\rmk - g^2\sum_{j=1}^{N-1}  |\sqrt{2}\Phi_j Q_{j+1}-vQ_j|^2 \rkk  \notag\\
&=\int\dd^4 x\lkk  \sum_{j=1}^N \lmk (D_\mu Q_j)^* D^\mu Q_j-M_Q^2 |Q_j|^2\rmk - g^2v^2\sum_{j=1}^{N-1}  |\tilde{\Phi}_j Q_{j+1}-Q_j|^2 \rkk \, ,
\end{align}
with $D_\mu Q_j = \del_\mu Q_j + iA_{j,\mu}Q_j$. The last term corresponds to the covariant derivative term in the fifth direction as in the fermion case, and the 5D bulk action can be reproduced under the correspondence,
\begin{align}
	a^{1/2} Q(x,y_j) \longleftrightarrow Q_j(x)\, . \label{scalar_correspondence}
\end{align}
Expanding $\tilde{\Phi}_j$, we further obtain
\begin{align}
    S_D^s & = \int \dd^4 x \lkk \sum_{j=1}^N  (D_\mu Q_j)^* D^\mu Q_j - \lmk M_Q^2+ g^2v^2\rmk\lmk |Q_1|^2+|Q_N|^2\rmk -\sum_{j=2}^{N-1}\lmk M_Q^2+2 g^2v^2\rmk|Q_j|^2\right. \notag\\
	& \left. + g^2 v^2 \sum_{j=1}^{N-1} \lmk Q_j^*e^{i\chi_j/v}Q_{j+1} + Q_j e^{-i\chi_j/v}Q_{j+1}^*\rmk  \rkk \, . \label{deconst_scalar}
\end{align}
We can remove all the Goldstone mode dependence from the scalar potential by field redefinition,
\begin{align}
	Q_2\rightarrow e^{i\chi_1/v} Q_2 \, , \quad Q_3\rightarrow e^{i(\chi_1+\chi_2)/v} Q_3 \, , \quad \cdots\, , \quad Q_N\rightarrow e^{i \sum_j \chi_j/v} Q_N \, , \label{Q_redefinition}
\end{align}
leaving the derivative couplings induced from the scalar kinetic terms.
Then, we have the mass terms,
\begin{align}
	\mathcal{L}_{sm} &=  -\sum_{j=1}^N M_Q^2 |Q_j|^2 - g^2v^2\sum_{j=1}^{N-1}  |e^{i\chi_j/v}Q_{j+1}-Q_j|^2 \notag\\
	&=- \lmk M_Q^2+ g^2v^2\rmk\lmk |Q_1|^2+|Q_N|^2\rmk -\sum_{j=2}^{N-1}\lmk M_Q^2+2 g^2v^2\rmk|Q_j|^2 +g^2v^2 \sum_{j=1}^{N-1}
	\lmk Q_j^* Q_{j+1} + \mathrm{h.c.} \rmk \notag\\
    &= - \sum_{i,j} Q_i^* \lmk \mathcal{M}_Q^2\rmk_{ij} Q_j \, ,
\end{align}
with the mass matrix given by
\begin{align}
    \mathcal{M}_Q^2 = M_Q^2 \mathbf{1}_N + g^2v^2 \Delta_N^{\rm N}\, ,
\end{align}
where $\mathbf{1}_N$ denotes the $N$-dimensional identity matrix.
Therefore, $Q_{j}$ have the same eigenvectors as $G_{j,\mu}$, and the spectrum can be written as
\begin{align}
    m_k^2 &= M_Q^2+4g^2 v^2 \sin^2\lmk \frac{k\pi}{2N}\rmk\, , \label{eigenmass_scalar}\\[1ex]
    Q^{(k)} &= \sqrt{\frac{c_k}{N}}\sum_{j=1}^N \cos\lmk\frac{\lmk j-\half\rmk k\pi}{N}\rmk Q_j\, ,\quad \text{with} \,\,\, k=0,1,\cdots, N-1\, . \label{eigenstate_scalar}
\end{align}
Again, the 5D KK spectrum~\eqref{scalar_KK_spectrum} is reproduced when $N$ is taken to infinity.

\section{Gauged WZW Term}
\label{sec:WZW}
The axion-gluon coupling in the deconstruction picture is given by the counterpart of 5D Chern-Simons term, the so-called  gauged Wess-Zumino-Witten (WZW) term. While it is not straightforward to derive this term by discretizing the CS term directly, 
we consider the integration of the introduced charged fermions $\Psi_i(x)$ to find the explicit form of the WZW term, as the CS term can be also generated by integrating out a charged fermion in the bulk. 

\subsection{Construction and Validation}

For simplicity, we take $M_\Psi'=0$ and the fermion interactions are left with hopping terms
$-gv\sum_{i=1}^{N-1}\lmk \psi_i\tilde{\Phi}_i\tilde{\Sigma}_i\eta_{i+1} + {\rm h.c.}\rmk$.
Let us focus on one unit of hopping interaction, which includes Weyl spinors $\psi_i,\eta_{i+1}$ and link fields $\Phi_i, \Sigma_i$,
\begin{align}
	\mathcal{L}_{{\rm unit},i}= \eta_{i+1}^\dag i\overline{\sigma}^\mu D_\mu\eta_{i+1}+\psi_i^\dag i\overline{\sigma}^\mu D^*_\mu\psi_i-gv\lmk \psi_i\tilde{\Phi}_i\tilde{\Sigma}_i\eta_{i+1} + {\rm h.c.}\rmk\, .
\end{align}
The structure of this Lagrangian is analogous to the model discussed in Ref.~\cite{Manohar:1984uq}, where the left- and right-handed flavor symmetries are gauged and connected by a Goldstone field. In the present setup, the lattice sites $i$ and $i+1$ play the roles of the left- and right-handed flavor indices, and the corresponding symmetry is identified as $U(1)\times SU(3)$.
Because the gauge groups at sites $i$ and $i+1$ each couple to only a single chiral Weyl fermion, the fermion content is anomalous. In particular, the theory exhibits three types of gauge anomalies: the cubic $U(1)^3$ anomaly, the mixed $U(1)$--$SU(3)^2$ anomaly, and the cubic $SU(3)^3$ anomaly, whose explicit forms are given below.

First we clarify the change of some conventions and notations within  the current section such that the equations are much simplified.
We take the convention using the differential forms, for example, the gauge potentials and field strengths are denoted as $A_i\equiv A_{i,\mu}\dd x^\mu,~\mathcal{F}_i\equiv \half \mathcal{F}_{i,\mu\nu}\dd x^\mu\wedge \dd x^\nu,~G_i\equiv G_{i,\mu} \dd x^\mu$ and $\mathcal{G}_i \equiv \half \mathcal{G}_{i,\mu\nu}\dd x^\mu\wedge \dd x^\nu$.
Besides, we turn to the convention where the generators are anti-Hermitian by absorbing the imaginary unit.
For example, the link fields are now written as $\tilde{\Phi}_i = e^{\chi_i/v}, \tilde{\Sigma}_i = e^{\pi_i/v_3}$.
Also due to this convention, the infinitesimal gauge transformations now look like
\begin{align}
    A_i\rightarrow A_i- \dd \alpha_{i}\, , ~A_{i+1} \rightarrow A_{i+1}- \dd \alpha_{i+1}\, , ~\tilde{\Phi}_i\rightarrow \tilde{\Phi}_i (1+\alpha_{i}-\alpha_{i+1})\, , \notag \\[1ex]
    \psi_i\rightarrow \psi_i(1-\alpha_{i}),~\eta_{i+1}\rightarrow \eta_{i+1}(1+\alpha_{i+1})\, , \\[1ex]
    G_i\rightarrow G_i - \dd \epsilon_i + [\epsilon_i, G_i] \, ,~ G_{i+1}\rightarrow G_{i+1} - \dd \epsilon_{i+1} + [\epsilon_{i+1}, G_{i+1}]\, , \notag \\[1ex] 
    \tilde\Sigma_i\rightarrow \tilde\Sigma_i + \epsilon_i\tilde \Sigma_i - \tilde \Sigma_i\epsilon_{i+1} \, ,~\psi_i\rightarrow \psi_i - \psi_i\epsilon_i\, ,~\eta_{i+1}\rightarrow \eta_{i+1} + \epsilon_{i+1}\eta_{i+1}\, ,
\end{align}
where $\alpha_{i}, \alpha_{i+1}, \epsilon_{i}\equiv \epsilon_{i}^a T^a, \epsilon_{i+1}\equiv \epsilon_{i+1}^a T^a$ are the gauge transformation parameters of the $U(1)_{i}, U(1)_{i+1}, SU(3)_{i}, SU(3)_{i+1}$ groups, respectively. 

We can learn how the action transforms under the gauge transformations from the charge assignment of the fermions.
Under the $U(1)_{i}, U(1)_{i+1}$ transformations, the action varies as
\begin{align}
    \delta_1 S_{\mathrm{unit},i} = \int_{R^4} -\frac{3}{32\pi^2} \lkk \alpha_i \mathcal{F}_i^2 -\alpha_{i+1} \mathcal{F}_{i+1}^2 \rkk -\frac{1}{16\pi^2} \lkk \alpha_i\tr\lmk \mathcal{G}_i^2\rmk -\alpha_{i+1} \tr \lmk \mathcal{G}^2_{i+1}\rmk\rkk \, , \label{anomaly_variation_u1}
\end{align}
where the notation $\mathcal{F}_i^2, \mathcal{G}_i^2$ refers to the wedge product of the field strengths.
Here the first term corresponds to the $U(1)_i^3,~U(1)_{i+1}^3$ anomalies and the factor of $3$ is due to the fermions as $SU(3)$ triplets while the second term comes from the mixed $U(1)_i-SU(3)_i^2,~U(1)_{i+1}-SU(3)_{i+1}^2$ anomalies.
Under the $SU(3)_{i}, SU(3)_{i+1}$ transformations, the action varies as
\begin{align}
    \delta_3 S_{\mathrm{unit},i} = \int_{R^4} -\frac{1}{24\pi^2} \tr\lkk \epsilon_i\lmk \dd G_i \dd G_i+\half \dd(G_i^3)\rmk -\epsilon_{i+1}\lmk \dd G_{i+1} \dd G_{i+1}+\half \dd(G_{i+1}^3)\rmk\rkk \, . 
\end{align}
In the low-energy effective theory where the massive fermions are integrated out, the information of the anomalies should be kept for consistency.
In fact, these anomalies are manifest in the gauged WZW term, which has been derived for the case of $SU(3)^3$ in Ref.~\cite{Manohar:1984uq}, and Ref.~\cite{Skiba:2002nx} showed that it can reproduce the 5D CS term consisting of only the $SU(3)$ gauge field.
Here we extend the result to the case of the mixed $U(1)-SU(3)^2$ anomaly and propose the corresponding WZW term,
\begin{align}
	\mathcal{L}_{{\rm WZW},i}^{\rm mix} &= \frac{1}{16\pi^2}\lkk\lmk A_i+ A_{i+1}+\tilde{\Phi}_i^* \dd \tilde{\Phi}_i\rmk \tr\lmk G_{i+1}\dd G_{i+1} +\frac{2}{3} G_{i+1}^3\rmk -\frac{1}{3}A_i\tr\lmk\tilde{\Sigma}_i\dd\tilde{\Sigma}_i^\dag\rmk^3 \right. \notag\\
	& \left. + \dd A_i \tr\lmk \tilde{\Sigma}_i^\dag \dd \tilde{\Sigma}_i G_{i+1} +G_i\tilde{\Sigma}_i G_{i+1}\tilde{\Sigma}_i^\dag +G_i\tilde{\Sigma}_i \dd\tilde{\Sigma}_i^\dag\rmk  -{\rm p.c.}\rkk \, . \label{unit_WZW}
\end{align}
Here, $\rm p.c.$ represents the parity conjugate under which $A_i \leftrightarrow A_{i+1},~ G_i \leftrightarrow G_{i+1} ,~ \tilde{\Phi}_i\leftrightarrow \tilde{\Phi}_i^*,$ and $\tilde{\Sigma}_i\leftrightarrow \tilde{\Sigma}_i^\dag$. 
Under the $U(1)_{i}, U(1)_{i+1}$ transformations, Eq.~\eqref{unit_WZW} transforms as
\begin{align}
	\delta \mathcal{L}_{{\rm WZW},i}^{\rm mix} &=\frac{1}{16\pi^2}\lkk \dd\alpha_i \tr\lmk \omega_{3,i}\rmk- \dd\alpha_{i+1} \tr\lmk \omega_{3,i+1}\rmk\rkk \nonumber \\[1ex]
	&= \frac{1}{16\pi^2}\lkk -\alpha_i \tr\lmk \mathcal{G}_i^2\rmk + \alpha_{i+1} \tr\lmk \mathcal{G}_{i+1}^2\rmk \rkk \label{WZW_anomaly}\, ,
\end{align}
where $\omega_{3,i}=G_{i}\dd G_{i} +\frac{2}{3} G_{i}^3$ is a differential 3-form whose exterior derivative is $\dd \omega_{3,i} = G_i^2$ and Eq.~\eqref{WZW_anomaly} is obtained by integration by parts. 
We can also verify that our WZW term is invariant under the $SU(3)_i,SU(3)_{i+1}$ gauge transformations. Therefore, it correctly reproduces the $U(1)_i-SU(3)_i^2$ and $U(1)_{i+1}-SU(3)_{i+1}^2$ anomalies.

Next, we further validate the WZW term we have proposed by reproducing the 5D CS term in the continuum limit of the deconstruction setup.
Going back to the whole lattice, the total WZW term is obtained by summing over all the units,\footnote{Here we formally introduce the $A_{1},A_{N}$ gauge fields while the Dirichlet boundary condition will be discussed in the next subsection.}
\begin{align}
	S_{\rm WZW}^{\rm mix} &= \sum_{i=1}^{N-1} \int_{R^4} \frac{1}{16\pi^2}\lkk\lmk A_i+ A_{i+1}+\tilde{\Phi}_i^* \dd \tilde{\Phi}_i\rmk \tr\lmk G_{i+1}\dd G_{i+1} +\frac{2}{3} G_{i+1}^3\rmk \right. \notag\\ 
	& \left. -\frac{1}{3}A_i\tr\lmk\tilde{\Sigma}_i\dd\tilde{\Sigma}_i^\dag\rmk^3  +\dd A_i \tr\lmk \tilde{\Sigma}_i^\dag \dd \tilde{\Sigma}_i G_{i+1} +G_i\tilde{\Sigma}_i G_{i+1}\tilde{\Sigma}_i^\dag +G_i\tilde{\Sigma}_i \dd\tilde{\Sigma}_i^\dag\rmk  -{\rm p.c.}\rkk  \, . \label{WZW_action}
\end{align}
According to Eq.~\eqref{WZW_anomaly}, under the collective $U(1)_i$ gauge transformations, the variations of $S_{\rm WZW}^{\rm mix}$ at sites $2\leq i \leq N-2$ are canceled by contributions from the adjacent sites, leaving only the boundary variations,
\begin{align}
	\delta S_{\rm WZW}^{\rm mix} = \frac{1}{16\pi^2}\lkk-\alpha_1\tr\lmk \mathcal{G}_1^2\rmk +\alpha_N\tr\lmk \mathcal{G}_N^2\rmk\rkk\, ,\label{delta_WZW}
\end{align}
which is formally consistent with the variation of the 5D Chern-Simons term in Eq.~\eqref{delta_CS}. 
Furthermore, Eq.~\eqref{WZW_action} exactly reproduces the CS term with $\kappa_{\rm CS}=1$ when we switch back to the 5D picture by replacing
$A_i\leftrightarrow A\, , A_{i+1}\leftrightarrow A+a\del_5 A\, , \tilde{\Phi}_i \leftrightarrow 1+aA_5\, , G_i\leftrightarrow G\, , G_{i+1}\leftrightarrow G+a\del_5 G\, ,\tilde{\Sigma}_i\leftrightarrow 1+a G_5 $ and $\sum_i a\leftrightarrow \int dy$,
\begin{align}
	S_{\rm WZW}^{\rm mix}\rightarrow -\int_{R^4\times I} \frac{1}{8\pi^2} A\tr\lmk \mathcal{G}^2\rmk=-\int\dd^4x\int_0^{\pi R}\dd y \frac{1}{32\pi^2}\epsilon^{MNPQR} A_M\tr \lmk \mathcal{G}_{NP} \mathcal{G}_{QR}\rmk\, , 
\end{align}
where $A\equiv A_M(x,y)\dd x^M, G\equiv G_M(x,y)\dd x^M, \mathcal{G}\equiv \frac{1}{2}\mathcal{G}_{MN}\dd x^M\wedge \dd x^N$ are differential forms defined in 5D. 
The detailed calculation for reconstructing the CS term is shown in Appendix~\ref{sec:appendix_WZW_CS}. 

\subsection{Alternative Approach for Dirichlet BC}
There exists one subtle issue that in our deconstruction setup, $U(1)_1$ and $U(1)_N$ gauge fields are not present due to the Dirichlet boundary condition. 
Therefore, the first and last units are not complete in the expression~\eqref{WZW_action}.
In order to address this issue, we can consider an alternative approach to realize the Dirichlet boundary condition, which introduces both the gauge fields $A_1(x),A_N(x)$ and two additional Higgs fields $H_1, H_N$ charged under $U(1)_1, U(1)_N$ respectively. 
We assume some potential for the Higgs fields so that $H_1,H_N$ both get a VEV which is much larger than that of the link fields $H_1\sim \frac{v'}{\sqrt{2}}e^{i\phi_1/v'}\, ,H_N\sim \frac{v'}{\sqrt{2}}e^{i\phi_N/v'}\, ,$ where $v'\gg v$.\footnote{We assume $H_1,H_N$ share the same VEV just for simplicity, and different VEVs will not affect the discussion.} 
Now, the link fields $\Phi_1,\Phi_{N-1}$ also have charge $+1,-1$ under $U(1)_1, U(1)_N$, respectively.
 As a result, the mass term~\eqref{A_mass} is modified into
 \begin{align}
 \mathcal{L}&\supset-\frac{1}{2}\lkk \lmk g v' \hat A_{1,\mu}+\del_\mu\phi_1 \rmk^2+\sum_{j=1}^{N-1}\lmk gv \hat A_{j+1,\mu}-gv \hat A_{j,\mu}+\del_\mu\chi_j\rmk^2+ \lmk gv' \hat A_{N,\mu}+\del_\mu\phi_{N}\rmk^2\rkk\notag\\
&\supset \frac{1}{2}g^2 v^2\lkk \lmk \frac{v'}{v}\rmk^2 \lmk \hat A_{1,\mu}^2+A_{N,\mu}^2 \rmk + \sum_{j=1}^{N-1}\lmk \hat A_{j+1,\mu}- \hat A_{j,\mu}\rmk^2 \rkk \, . \label{modified_mass_term}
\end{align}
We can expect that since $v'\gg v$, $\hat A_{1,\mu}, \hat A_{N,\mu}$ are much heavier than the other gauge fields and almost decouple.
In such a decoupling limit, we can pull $\hat A_{1,\mu}, \hat A_{N,\mu}$ out of the second term in the last equality.
Then we get the same mass term for $\hat A_{j,\mu},~j=2,\cdots, N-1$ as Eq.~\eqref{A_mass} and the masses of these two heavy gauge fields are just
\begin{align}
    m_{\rm heavy}^2 \simeq g^2{v'}^2(1+\frac{v^2}{{v'}^2}) \, .
\end{align}
In Appendix~\ref{sec:appendix_precise_spectrum}, we show the precise calculation of the mass spectrum in such scenario, where the mass squared only gets corrections at $O(\frac{v^4}{{v'}^4})$.

This approach can be understood in the 5D picture as follows. 
First we consider a $U(1)$ gauge theory with the Neumann boundary conditions imposed on $A_\mu$, corresponding to our deconstruction setup without the extra Higgs fields.
Then we introduce boundary localized $U(1)$ Higgs fields $H_{IR}(x,y=0),~H_{UV}(x,y=\pi R)$ with a VEV $v'$, so that the boundary conditions are modified into
\begin{align}
	\del_y A_\mu \mp v' A_\mu|_{y=0,\pi R}=0\, .
\end{align}
Since $v'\gg v$ is much larger than the energy scale of the 5D gauge theory, it can be considered as infinity.
Then the Dirichlet boundary conditions are reproduced dynamically.
The similar discussion can be found in Ref.~\cite{Csaki:2005tasi}.
In this way, all the units of the WZW term are complete and we can consistently get Eq.~\eqref{WZW_action}.
Since the difference between two approaches is not relevant for later discussions, for simplicity, we just take the conventional setup where $A_1,A_N$ are removed.
Consequently, the $U(1)$ variation of the WZW term~\eqref{delta_WZW} vanishes just as that of the 5D CS term~\eqref{delta_CS} after imposing the Dirichlet boundary conditions. 

\subsection{Axion-Gluon Coupling}
In the low-energy effective theory after integrating out the massive fermions, the coupling between the $U(1)$ Goldstones $\chi_j$ and the $SU(3)$ gauge fields $G_j$ is encoded in the WZW term,
\begin{align}
	\mathcal{L}_{\chi GG}&= \sum_j \frac{1}{16\pi^2}\lkk\tilde{\Phi}_j^* \dd \tilde{\Phi}_j \tr\lmk\omega_{3,j+1}\rmk - \tilde{\Phi}_j \dd \tilde{\Phi}_j^* \tr\lmk\omega_{3,j}\rmk \rkk\notag\\
	&=\frac{1}{16\pi^2} \sum_i\tilde{\Phi}_j^* \dd \tilde{\Phi}_j \tr\lmk\omega_{3,j+1}+\omega_{3,j}\rmk= \frac{1}{16\pi^2} \sum_j\frac{\dd \chi_j}{v} \tr\lmk\omega_{3,j+1}+\omega_{3,j}\rmk \notag\\
	&=-\frac{1}{32\pi^2} \sum_i \frac{\chi_j}{v} \tr\lmk \mathcal{G}_{j,\mu\nu} \widetilde{\mathcal{G}_j}^{\mu\nu} + \mathcal{G}_{j+1,\mu\nu} \widetilde{\mathcal{G}_{j+1}}^{\mu\nu}\rmk \, ,
	\label{chi_GG}
\end{align}
where in the last equality, we perform the integration by parts and use the relation,
\begin{align}
\int\dd^4 x \tr\lmk \mathcal{G}_{j,\mu\nu}\widetilde{\mathcal{G}_{j}}^{\mu\nu}\rmk = \int\dd^4 x \half\epsilon^{\mu\nu\rho\sigma} \tr \lmk \mathcal{G}_{j,\mu\nu} \mathcal{G}_{j,\rho\sigma} \rmk = \frac{1}{2} \int 4\tr\lmk\mathcal{G}_j^2\rmk =-2\int \dd \tr(\omega_{3,j}) \, .
\end{align}
Switching to the mass eigenstates of the Goldstone modes,
\begin{align}
	\chi_j &= \frac{1}{\sqrt{N-1}} a(x) +\sqrt{\frac{2}{N-1}}\sum_{k=1}^{N-2}\cos\lmk \frac{(j-\half)k\pi}{N-1}\rmk \chi^{(k)}\, ,
\end{align}
we get the axion-gluon coupling as
\begin{align}
	\mathcal{L}_{a GG}&= -\frac{1}{32\pi^2} \frac{a(x)}{\sqrt{N-1}v} \sum_{i=1}^{N-1}  \tr\lmk \mathcal{G}_{i,\mu\nu} \widetilde{\mathcal{G}_i}^{\mu\nu} + \mathcal{G}_{i+1,\mu\nu} \widetilde{\mathcal{G}_{i+1}}^{\mu\nu} \rmk \\
	&=-\frac{1}{32\pi^2} \frac{a(x)}{\sqrt{N-1}v} \tr\lmk  \mathcal{G}_{1,\mu\nu} \widetilde{\mathcal{G}_1}^{\mu\nu} + 2\sum_{i=2}^{N-1}\mathcal{G}_{i,\mu\nu} \widetilde{\mathcal{G}_i}^{\mu\nu} +  \mathcal{G}_{N,\mu\nu} \widetilde{\mathcal{G}_{N}}^{\mu\nu} \rmk \, . \label{aGG_lattice}
\end{align}
Again, in order to obtain the axion coupling to the massless gluon, we replace $G_i(x)$ with their mass eigenstates,
\begin{align}
	G_{i,\mu} &= G^{(0)}_\mu + \sqrt{2}\sum_{k=1}^{N-1}\cos\lmk \frac{(j-\half)k\pi}{N-1}\rmk G^{(k)}_\mu\, , \label{gauge_mass_eigenstate}
\end{align}
and keep only the zero mode $G_\mu^{(0)}(x)$ explicitly,
\begin{align}
	\mathcal{L}_{a GG} = -\frac{1}{16\pi^2} \frac{\sqrt{N-1} a(x)}{v} \tr\lmk \mathcal{G}^{(0)}_{\mu\nu} \widetilde{\mathcal{G}^{(0)}}^{\mu\nu}  \rmk +\cdots \, ,\label{aGG_mass}
\end{align}
where $\cdots$ contains the coupling between the axion and the massive modes of the $SU(3)$ gauge bosons.
In fact, Eq.~\eqref{gauge_mass_eigenstate} cannot be applied to the field strengths due to their non-linear feature, where different massive modes mix with each other in the cubic and quartic terms. 
Therefore, the couplings between the axion and the massive modes are quite complicated and cannot be written in the form like $a\mathcal{G}\widetilde{\mathcal{G}}$. 
We are only interested in the axion coupling with the massless mode, from which we can read the decay constant, 
\begin{align}
	f_a=\frac{v}{\sqrt{N-1}}=\frac{\sqrt{v/g}}{\sqrt{(N-1)(gv)^{-1}}} \leftrightarrow \frac{1}{g_5\sqrt{\pi R}}\equiv f_{a,5}\, .
\end{align}
Here, we implicitly take the continuum limit so that $N-1\simeq N$, and it is shown that the axion decay constants in two pictures are consistent. We now have the definition for the dimensionless axion field,
\begin{align}
	\theta(x) \equiv \frac{a(x)}{f_a} = \sum_{j=1}^{N-1} \frac{\chi_j(x)}{v}~. \label{deconst_theta}
\end{align}

\section{Axion Potential and the Quality Problem
\label{sec:quality}}
As mentioned in Sec.~\ref{sec:deconstruction}, the axion in the deconstruction theory is identified as $a(x)\equiv \frac{1}{\sqrt{N-1}}\sum_j\chi_j(x)$.
Through its coupling with the massless $SU(3)$ gauge field, which we identify with the QCD gluon, the axion can dynamically solve the strong CP problem.
However, any interaction that violates the axion shift symmetry can lead to a shift from the CP-conserving vacuum and spoil the solution, known as the axion quality problem.
The extra-dimensional axion is considered as one realization of the high-quality axion because the axion shift symmetry defined as $\int \dd y A_5\rightarrow \int\dd y A_5 + c$ is protected by the bulk $U(1)$ gauge symmetry.
The sources of shift symmetry breaking are limited to the effects that are non-local in the fifth dimension~\cite{Reece:2025thc,Choi:2026kxu}:
\begin{itemize}
    \item Gauge instantons: an extra gauge field may couple to the $U(1)$ gauge field through the Chern-Simons coupling and its instanton effect would generate an additional potential to the axion.
    \item Bulk charged matter: some matter fields charged under the $U(1)$ gauge group also contribute to the axion potential, which can be interpreted as arising from their propagation across the fifth dimension.
\end{itemize}

In this section, we will investigate the axion quality problem in the deconstruction setup.
The axion shift symmetry is identified with a global $U(1)$ symmetry under which all link fields $\Phi_j$ are identically charged. In order for the Yukawa term $\psi_j\Phi_j\eta_{j+1}$ and the scalar interaction $Q_j^* \Phi_j Q_{j+1}$ to preserve this symmetry, the charge assignment is set as in Tab.~\ref{tab:axion_shift_charge}.
\begin{table}[t!]
    \centering
    \begin{tabular}{c|c|c|c|c}
         & $\Phi_j$ & $\psi_j$ & $\eta_{j}$ & $Q_j$\\
        $U(1)_{\rm shift}$ & $+1$ & $-\frac{1}{2}$ & $-\frac{1}{2}$ & $j$ 
    \end{tabular}
    \caption{Charge assignments under the axion shift symmetry.}
    \label{tab:axion_shift_charge}
\end{table}
Similar to the discussion in the extra dimension picture, the $U(1)_{\rm shift}$-breaking sources are highly constrained by the $\prod_{j=2}^{N-1}U(1)_j$ gauge symmetries.
Firstly, we notice that we can directly write down a higher-dimensional $U(1)_{\rm shift}$-breaking term using only the link fields which preserve all the gauge symmetries,
\begin{align}
    V_{\cancel{\rm shift}} = \frac{\kappa}{\Mpl^{N-5}}\prod_{j=1}^{N-1}\Phi_j +\mathrm{h.c.} = 2 |\kappa| \Mpl^4 \lmk\frac{v}{\sqrt{2}\Mpl}\rmk^{N-1} \cos\lmk \frac{a}{f_a}+\delta_\kappa\rmk \, , \label{shift_breaking_term}
\end{align}
where $\delta_\kappa=\arg \kappa$ and the cut-off scale is chosen as the Planck scale assuming that the symmetry breaking is caused by quantum gravity effects as commonly considered in the literature.
The collective $U(1)_j$ gauge invariance requires the appearance of the product $\prod_j\Phi_j$ and leads to the suppression factor.
Obviously, for $v<\Mpl$ and $N \gg 1$, Eq.~\eqref{shift_breaking_term} is negligible and safe for the axion quality.

We will discuss the two non-local effects mentioned above in terms of the deconstruction setup and show the correspondence between the two pictures. 
First, we consider the axion potential originating from the instanton effect of some gauge sector with the gauge fields scattering across the lattice and coupled to the axion through the gauged WZW term introduced in Sec.~\ref{sec:WZW}.
As discussed in Refs.~\cite{Poppitz:2002ac,Gherghetta:2020keg}, such instanton effects can be UV dominated due to the enhancement from the tower of massive modes.
Therefore, we will focus on the small instanton effects.
Second, we study the effect of the bulk charged matter, taking the scalar fields $Q_j(x)$ as an example.
\subsection{Small Instanton Effects}
For convenience, we take the $SU(3)$ sector introduced in Sec.~\ref{sec:deconstruction} as an example,\footnote{Since we have not introduced any Standard Model particles charged under the $SU(3)$ yet, we can temporarily forget its role as the gluon.} which can be easily generalized to the case of $SU(N)$.
According to the previous discussion, we have the axion coupling with the $SU(3)$ gauge fields~\eqref{aGG_lattice}, \eqref{aGG_mass}. 
In Eq.~\eqref{aGG_lattice}, especially, we see that the axion has the typical topological interactions with the $SU(3)$ gauge fields at all the lattice sites and they all contribute to the axion potential through instanton effects. 
One may consider different instanton configurations by randomly assigning the instanton solutions to $G_{j,\mu}(x),~j=1,\cdots,N$. 
One notation for these configurations utilizes an $N$-tuple where the $j$-th number is the instanton number of the $SU(3)_j$ gauge field.
We constrain ourselves to the cases of the single instanton solutions, and the possible configurations include $(1,0,\cdots,0),\cdots,(0,\cdots,0,1),\cdots,(1,1,\cdots,1)$. The $j$-th number being $1$ means $G_{j,\mu}$ has a non-trivial classical configuration given by the BPST instanton solution~\cite{Belavin:1975fg,tHooft:1976snw},
\begin{align}
    G_{j,\mu}(x) = 2\rho_j^2 \frac{\bar \eta^a_{\mu\nu} (x-x_{0,j})_\nu}{(x-x_{0,j})^2((x-x_{0,j})^2-\rho_j^2)^2} J^a_{\Omega_j} \equiv G_\mu^{\rm BPST}(x;x_{0,j},\rho_j,\Omega_j) \, ,
\end{align}
where $x_{0,j},\rho_j$ denote the center position and the size of the instanton, respectively, $\bar{\eta}^{a}_{\mu\nu}$ is the anti-self-dual ’t Hooft symbol, $J^a_{\Omega_j}$ are the generators of the $SU(2)$ subgroup and $\Omega_j$ represents its orientation inside the $SU(3)_j$.
For example, the minimal embedding is given by putting the $SU(2)$ generators on the top-left corner of the $3\times 3$ matrices which generate the fundamental representation of $SU(3)_j$.

The discussion of the instanton effects highly depends on the size of the instanton. 
We first consider a very small instanton with $\rho_j\ll v_3^{-1}$.
In this region, we can approximately take $v_3=0$ and thus all the instantons are independent of each other.
The $j$-th instanton amplitude with topological charge $\Delta Q =1$ is proportional to
\begin{align}
     W_{I,j} \equiv \frac{Z_{\Delta Q=1,j}}{Z_{0,j}} \propto \exp\lmk -8\pi^2/g^2(\rho_j^{-1})\rmk \, ,
\end{align}
and the total single instanton amplitude is given by
\begin{align}
    W_{I,\rm tot} = \prod_{j=1}^N \lmk W_{I,j}\rmk^{n_j} \, ,
\end{align}
where $n_j=0,1$ is the $j$-th instanton number which is determined by the configuration that we consider.
Since all the $SU(3)_j$ gauge groups are asymptotically free with only the link fields considered as the charged matter, the instanton amplitudes are extremely suppressed for small $\rho_j$.

As $\rho_j$ approaches $v_3$, the instantons start to feel the effect of Higgsing.
There remains only one precise solution to the Euclidean EoM, the zero-mode instanton $(1,1,\cdots,1)_{\rm zero}$ defined as
\begin{align}
    G_{j,\mu}(x) = G_\mu^{\rm BPST}(x;x_0,\rho,\Omega)\,,~~~j=1,\cdots,N\, ,
\end{align}
where the instanton backgrounds at all sites are aligned due to the mass mixing terms $\frac{1}{2}v_3^2\lmk G_{j+1,\mu}-G_{j,\mu}\rmk^2$~\cite{Poppitz:2002ac}.
Therefore, we suppress the subscript $j$ in the instanton parameters $x_0,\rho,\Omega$.
As implied by the name, this configuration is equivalent to the instanton background of the zero mode gauge field $G^{(0)}_\mu$.
This is consistent with the 5D picture since the 5D instanton solution is given by uplifting the 4D BPST instanton solution to simply wrap around the fifth dimension~\cite{Gherghetta:2020keg},
\begin{align}
	G_\mu(x,y) = G_\mu^{\rm BPST}(x)\, , ~ G_5(x,y) = 0 ~\longleftrightarrow ~G_{j,\mu}(x) = G_\mu^{\rm BPST}(x)\,,~~j=1,\cdots,N\, .
\end{align}
Here, we suppress the dependence of the instanton solution on the parameters $x_0,\rho,\Omega$ just for simplicity.

Other instanton configurations, including more general $(1,1,\cdots,1)$-instantons with different $x_{0,j},\rho_j,\Omega_j$ and the ``fractional'' instantons\footnote{Here the name of the ``fractional" instanton comes from the comparison with the ``full" $(1,1,\cdots,1)$ instanton and is different from the fractional instantons in other contexts such as $SU(N)/Z_N$.} such as $(1,0,\cdots,0)$-instanton, still remain as the constrained instanton solutions~\cite{tHooft:1976snw,Affleck:1980mp} and potentially give sizable contributions to the axion potential.
It seems to be inconsistent with the 5D picture since these constrained instantons do not have the corresponding descriptions in 5D.
However, we show that, in the 5D regime $R^{-1} < \rho_j^{-1} <\Lambda_5 $, far below the VEV scale $v_3$ in a deconstruction setup that can successfully reproduce the 5D features, the constrained instanton contribution is highly exponentially suppressed.
In other words, the constrained instantons including the ``fractional'' ones would not affect the low-energy effective theory, namely, the 5D gauge theory, in this regime.

Below, our main interest is in the region $R^{-1} < \rho_j^{-1} <\Lambda_5 $ and we show explicitly the calculation of the zero-mode instanton effect as well as the reason that the constrained instanton effects are negligible, taking the ``fractional'' instanton as a typical example.

\subsubsection{The Zero-mode Instanton \label{zeroinstanton}}
First we consider the small instanton effect from the zero mode $G_\mu^{(0)}(x)$.
We work in the classical background,
\begin{align}
    G_{j,\mu}(x) = G_\mu^{\rm BPST}(x)\,,~~j=1,\cdots,N  ~~~\Longleftrightarrow~~~ G_\mu^{(0)}(x) = G_\mu^{\rm BPST}(x)\, , \label{zero_mode_bkg}
\end{align}
or equivalently, the $(1,1,\cdots,1)_{\rm zero}$ configuration. The corresponding gauge group is the unbroken diagonal group $SU(3)^{(0)}$ and under its infinitesimal transformation, the gauge fields in the lattice basis transform universally,
\begin{align}
	\Delta_\alpha G_{j,\mu} = \partial_\mu\alpha(x)+i\lkk \alpha(x),G_{j,\mu} \rkk .
\end{align}
According to Eq.~\eqref{eigenstate_su3}, the transformation laws for the massive gauge fields are given by
\begin{align}
	\Delta_\alpha G_\mu^{(k\geq1)} &= \frac{\sqrt{2}}{N} \sum_{j=1}^{N} \cos\lmk \frac{\lmk j-\half \rmk k\pi}{N}\rmk \Delta_\alpha G_{j,\mu}  \notag\\
	& = \frac{\sqrt{2}}{N} \del_\mu\alpha(x) \sum_{j=1}^{N} \cos\lmk \frac{\lmk j-\half \rmk k\pi}{N}\rmk + i \lkk \alpha(x), G_\mu^{(k)}\rkk = i \lkk \alpha(x), G_\mu^{(k)}\rkk \, ,
\end{align}
where the summation $\sum_{j=1}^{N} \cos\lmk \frac{\lmk j-\half \rmk k\pi}{N}\rmk$ vanishes unless $k=0$.
Hence, all the massive modes transform as the adjoint representation of $SU(3)^{(0)}$ and contribute to the zero-mode instanton amplitude.

We introduce, upon the background~\eqref{zero_mode_bkg}, the fluctuations $\delta G_{j,\mu}(x)$ and $\delta \pi_j(x)\equiv \delta \pi_j^a(x) T^a$, defined as
\begin{align}
	G_{j,\mu}(x) = G_\mu^{\rm BPST}(x) + \delta G_{j,\mu}(x) \, , \quad \Sigma_j(x) \simeq v_3 \lmk1+\frac{i\delta\pi_j(x)}{v_3}\rmk = v_3 + i\delta\pi_j(x)\, . \label{fluctuations}
\end{align}
Substituting Eq.~\eqref{fluctuations} into the deconstructed action~\eqref{deconst_SU3}, performing the Wick rotation and keeping up to quadratic terms in fluctuations, we find the Euclidean action,
 \begin{align}
S_D^{SU(3)} \simeq S_{\rm cl} -\int d^4x \tr \lkk \frac{1}{2g_c^2} \sum_{j=1}^{N} \delta G_{j,\mu}{\mathcal{M}_{A}'}^{\mu\nu}\delta G_{j,\nu} +\sum_{j=1}^{N-1}\left(D_\mu^{I}\delta\pi_j + v_3\left(\delta G_{j,\mu}-\delta G_{j+1,\mu}\right) \right)^2\rkk .
\end{align}
Here, the linear term disappears because $G_\mu^{\rm BRST}$ satisfies the EoM, and the higher-order terms can be neglected when we calculate the instanton effect at the 1-loop level. 
The classical action is given by
\begin{align}
	S^{\rm cl} = -N\frac{8\pi^2}{g_c^2} = -\frac{8\pi^2}{g_{c,D}^2}\, ,
\end{align}
and the quadratic operator before any gauge fixing is defined as
\begin{align}
	 {\mathcal{M}_A'}^{\mu\nu} \delta G_{j,\nu} = -(D^I)^2\delta G_{j}^\mu +D^{I,\mu} D^{I,\nu}\delta G_{j,\nu} -2i\lkk \mathcal{G}^{\mathrm{BPST},\mu\nu},\delta G_{j,\nu}\rkk \, ,
\end{align}
with the covariant derivative in the instanton background,
\begin{equation}
  D_\mu^{I}X=\partial_\mu X -i\lkk G_\mu^{\rm BPST},X\rkk.
\end{equation}

To simplify the quadratic operator and cancel the mixing between the fluctuations $\delta\pi_j$ and $\delta G_{j,\mu}$, we take the background gauge by introducing the gauge-fixing term,
\begin{align}
	 S_{\rm GF} = -\frac{1}{2\xi^{(0)} g_c^2} \int d^4x\sum_{j=1}^{N}{\rm Tr}{\cal F}_j^2 ~~~ \text{with} ~~~ {\cal F}_j=D_\mu^I\delta G_j^\mu - \xi^{(0)} v_3 g_c^2\left(\delta \pi_j-\delta \pi_{j-1}\right) \, ,
\end{align} 
as well as the Faddeev-Popov ghost action,
\begin{align}
	S_{\rm gh} = -\frac{1}{2 g_c^2}\int d^4x\,\sum_{j,l=1}^{N}\tr\lmk \bar c_j\lkk-\left(D^I\right)^2\delta_{j l}+\xi^{(0)} v_3^2(\Delta_N)_{j l} \rkk  c_l \rmk ,
\end{align}
where the matrix $\Delta^{\rm N}_N$ is defined in Eq.~\eqref{Delta_N}.
We can switch from the lattice basis to the mass basis by the expansion,
\begin{align}
      \delta G_{j,\mu}(x)&=\delta G_\mu^{(0)}(x)+\sqrt{2}\sum_{n=1}^{N-1}\cos\left[\frac{\left(j-\frac12\right)n\pi}{N}\right]\delta G_\mu^{(n)}(x)\, ,\\
  c_j(x)&=c^{(0)}(x)+\sqrt{2}\sum_{n=1}^{N-1}\cos\left[\frac{\left(j-\frac12\right)n\pi}{N}\right]c^{(n)}(x)\, ,\\
 \delta \pi_j(x)&=\sqrt{\frac{2}{N}}\sum_{n=1}^{N-1}\sin\left(\frac{jn\pi}{N}\right)\delta \pi^{(n)}(x)\, .
\end{align}
The full quadratic action upon the zero-mode (single) instanton background is then given by
\begin{align}
  S^I_{D} &\simeq-\frac{8\pi^2}{g_{c,D}^2}+\frac{1}{2g_{c,D}^2}\int d^4x\,{\rm Tr}\bigg[\delta G_\mu^{(0)} {\cal M}_{A}^{\mu\nu}\delta G_\nu^{(0)}+\bar c^{(0)}{\cal M}_{\rm gh}c^{(0)}\notag \\
  &+\sum_{n=1}^{N-1}\lmk \delta G_\mu^{(n)}\left(\mathcal{M}_{A}^{\mu\nu}+m_n^2\delta^{\mu\nu}\right)\delta G_\nu^{(n)} + \bar c^{(n)}\left(\mathcal{M}_{\rm gh}+m_n^2\right)c^{(n)}\rmk\bigg]  \notag\\
  &+\frac{1}{2}\int\dd^4 x \sum_{n=1}^{N-1}\delta\pi^{(n)}\left(\mathcal{M}_{\pi}+m_n^2\right)\delta\pi^{(n)} \, ,
\end{align}
where we take the Feynman-'t Hooft gauge $\xi^{(0)}=1$, the mass $m_n=\frac{n}{R}$, the quadratic operator of the gauge field is simplified as
\begin{align}
    \mathcal{M}_A^{\mu\nu} \delta G_{j,\nu} = -(D^I)^2\delta G_{j}^\mu  -2i\lkk \mathcal{G}^{\mathrm{BPST},\mu\nu},\delta G_{j,\nu}\rkk \, ,
\end{align}
and the quadratic operators of the ghosts and Goldstones are defined as
\begin{align}
    \mathcal{M}_{\rm gh}=\mathcal{M}_\pi=-\lmk D^I_\mu\rmk^2\, .
\end{align}

The single instanton amplitude can be evaluated by performing the path integral,
\begin{align}
    W_{I} &= \frac{Z_{\Delta Q=1}}{Z_0} = \frac{1}{Z_0} \int \prod_{n=0}^{N-1} D\delta G_\mu^{(n)} D c^{(n)} D \overline{c}^{(n)} \prod_{n=1}^{N-1} 
    D\delta\pi^{(n)}e^{-S_D^I} \notag \\
    & = \int \dd^4 x_0 \int \frac{\dd\rho}{\rho^5} C_3 \lmk \frac{2\pi}{\alpha_{c,D}}\rmk^6 e^{-S^{I}_{\rm eff}(\rho)} \, ,
\end{align}
where $\alpha_{c,D}=\frac{g_{c,D}^2}{4\pi}$,  the constant $C_3\simeq 1.5\times 10^{-3}$ and the calculation of the effective action $S_{\rm eff}^I(\rho)$ highly depends on the region of the instanton size.
The derivation of the 5D regime $\Lambda_5^{-1} < \rho < R$ has been presented in Refs.~\cite{Poppitz:2002ac,Gherghetta:2020keg} for both the $S^1$ and orbifold cases. The result is given by
\begin{align}
    W_I \supset \int \dd^4 x_0 \int_{\Lambda_5^{-1}}^R \frac{\dd\rho}{\rho^5} C_3 \lmk \frac{2\pi}{\alpha_{c,D}}\rmk^6 e^{-S^{I,5}_{\rm eff}(\rho)} \, ,
\end{align}
with the 5D-regime effective action calculated as
\begin{align}
    S^{I,5}_{\rm eff}(\rho)=\frac{2\pi}{\alpha_{c,D}(\pi gv/N)}- \frac{3N}{2\pi gv}\frac{1}{\rho} +\frac{47}{4}\ln\frac{N/\pi gv}{\rho} \, . \label{instanton_Seff}
\end{align}
The linear dependence of the effective action on $\rho^{-1}$ is typical of the 5D instanton and gives a significant enhancement of the instanton effect.
It comes from both the linear running behavior of the diagonal gauge coupling and the additional contributions from the massive gauge fields, which would give zero modes in the limit of vanishing masses, unlike the scalar fields whose effects are simply absorbed into the running of the gauge coupling.
The expression~\eqref{instanton_Seff} is valid for $\rho\lesssim R$ so that a large number of the massive modes are lighter than $\rho^{-1}$ and feel the instanton.
Since the linear behavior of $S_{\rm eff}^{I,5}(\rho)$ makes the instanton integral strongly UV dominated, the contribution from the $\rho \lesssim R$ region can be neglected.
Eq.~\eqref{instanton_Seff} also assumes that the masses of the massive gauge fields are in the same form as the KK masses, which is ensured by Eq.~\eqref{UV_completion_condition}.

The instanton amplitude $W_I$ is translated into the axion potential as follows. Taking the axion as a constant background field, it enters the Euclidean action by 
\begin{align}
i\frac{a}{f_a} \int \dd^4 x \frac{1}{16\pi^2} \tr\lmk \mathcal{G}^{(0)}_{\mu\nu}\widetilde{\mathcal{G}^{(0)}}^{\mu\nu} \rmk = i\frac{a}{f_a} \, ,
\end{align}
and therefore the effective action is modified into
\begin{align}
    S^I_{\rm eff}(\rho)\longrightarrow S^I_{\rm eff}(\rho) + i\frac{a}{f_a}
\end{align}
The instanton amplitude becomes
\begin{align}
    W_I[a] = \int\dd^4 x_0\, \mathcal{K} e^{-i a/f_a}\, ,
\end{align}
where 
\begin{align}
    \mathcal{K} \equiv  \int \frac{\dd\rho}{\rho^5} C_3 \lmk \frac{2\pi}{\alpha_{c,D}}\rmk^6 e^{-S^I_{\rm eff}(\rho)} \, .
\end{align}
Utilizing the dilute instanton gas approximation, the generating functional in the axion background is given by
\begin{align}
   Z[a] &= \sum_{n,\bar{n}=0}^{\infty} \frac{1}{n!\bar{n}!}(W_{I}[a])^{n}(W_{\bar{I}}[a])^{\bar{n}} = \exp\lmk W_{I}[a] + W_{\bar{I}}[a]\rmk  \notag \\
   &= \exp\lkk\int\dd^4 x_0 \, 2\mathcal{K} \cos\lmk\frac{a}{f_a}\rmk\rkk\, .
\end{align}
The axion potential is finally given by
\begin{align}
    V_{I}[a] = -\frac{1}{\mathcal{V}_4} \ln Z[a] = -2\mathcal{K} \cos\lmk\frac{a}{f_a}\rmk \, ,
\end{align}
where $\mathcal{V}_4$ is the volume of the 4-dimensional Euclidean space and is canceled by the integral $\int \dd^4 x_0$ since the integrand is independent of the center position of the instanton.
As we discussed above, the instanton integral of the 5D regime,
\begin{align}
    \mathcal{K}_5 \equiv \int_{\Lambda_5^{-1}}^R \frac{\dd\rho}{\rho^5} C_3 \lmk \frac{2\pi}{\alpha_{c,D}}\rmk^6 e^{-S^{I,5}_{\rm eff}(\rho)} \subset \mathcal{K} \, ,
\end{align}
is dominated by the small $\rho$ and thus can be roughly estimated as
\begin{align}
    \mathcal{K}_5 \simeq C_3\lmk\frac{2\pi}{\alpha_{c,D}(\Lambda_5)}\rmk^6 \Lambda_5^4 e^{-S_{\rm eff}^{I,5}(\Lambda_5^{-1})} \, .
\end{align}
It is shown in Ref.~\cite{Gherghetta:2020keg} that  due to the enhancement from the whole tower of the massive modes, $\mathcal{K}_5$ can already be larger than the ordinary axion potential generated by non-perturbative QCD effects.

\subsubsection{``Fractional" Instantons \label{sec:fracinstanton}}
Next we consider the ``fractional" instanton backgrounds and take the configuration,
\begin{align}
(\underbrace{0,\cdots,0}_{r},\underbrace{1,1,\cdots,1}_{s},\underbrace{0,\cdots,0}_{N-r-s}) \, ,
\end{align}
as an example. More explicitly, it is given by
\begin{align}
G_{j,\mu}(x) = 
\begin{cases}
    G_\mu^{\rm BPST}(x)\,,~&j=r+1,\cdots,r+s\, , \\
    0\, ,~& j=1,\cdots,r,r+s+1,\cdots, N\, .
\end{cases}
\end{align}
Here we assume the instantons at different sites are identical while more general configurations exist but are more suppressed as we explained above.
This can be seen as the instanton configuration of a sub-zero mode
when we focus on the sub-lattice including only the sites ranging from $r+1$ to $r+s$,
\begin{align}
    G^{(0)'}_\mu = \frac{1}{s}\sum_{j=r+1}^{r+s} G_{j,\mu}(x) \, ,
\end{align}
which corresponds to the diagonal subgroup $SU(3)^{(0)'}$ of $SU(3)_{r+1}\times\cdots\times SU(3)_{r+s}$.
In fact, this sub-zero mode becomes massive due to the  VEVs of the nearby link fields $\Sigma_r, \Sigma_{r+s}$, which transform as anti-fundamental and fundamental representations under $SU(3)^{(0)'}$.
This can be also seen by expanding the covariant derivatives of these two link fields,
\begin{align}
    D_\mu \Sigma_r &= \del_\mu \Sigma_r + i G_{r,\mu}\Sigma_r - i\Sigma_r G_{r+1,\mu} =\del_\mu \Sigma_r + i G_{r,\mu}\Sigma_r - i\Sigma_r G^{(0)'}_{\mu} + \cdots  \nonumber \\
    & \supset \del_\mu \Sigma_r - i\Sigma_r G^{(0)'}_{\mu} \, ,\\[1ex]
    D_\mu \Sigma_{r+s} &= \del_\mu \Sigma_{r+s} + i G_{r+s,\mu}\Sigma_{r+s} - i\Sigma_{r+s} G_{r+s+1,\mu} \notag\\
    &=\del_\mu \Sigma_{r+s} +i G^{(0)'}_{\mu}\Sigma_{r+s} + \cdots - i \Sigma_{r+s} G_{r+s+1,\mu}  \supset \del_\mu \Sigma_{r+s} +i G^{(0)'}_{\mu}\Sigma_{r+s}\, ,
\end{align}
where we expand $G_{r+1,\mu}$ and $G_{r+s,\mu}$ inside the sub-lattice as in Eq.~\eqref{eigenstate_su3}, $``\cdots"$ includes other modes in the expansion and the last expression includes only the sub-zero mode couplings that we are interested in.
The non-zero VEVs of $\Sigma_r,\Sigma_{r+s}$ will lead to mass terms for the sub-zero mode and make the non-trivial instanton configuration no longer the saddle point of the action.
However, we can still consider the so-called constrained instanton solution~\cite{Affleck:1980mp,Csaki:2019vte} by assigning non-trivial configurations to the link fields and requiring them to satisfy the EoM in the instanton background $D^2\lmk G^{(0)'}_\mu =G^{\rm BPST}_\mu\rmk \Sigma=0$, which leads to
\begin{align}
    \lmk\Sigma^{\rm cl}\rmk_{in}(x) = \begin{cases}
        \sqrt{\frac{x^2}{x^2+\rho^2}} \vev{\Sigma}_{in} \, ,\quad &i=1,2\, , \\
        \vev{\Sigma}_{in} \, ,\quad &i=3\, ,
    \end{cases} 
\end{align}
where $\Sigma_{in}$ represents $\lmk \Sigma_{r+1}\rmk _{in},~\lmk \Sigma_{r+s}^\dag\rmk _{in}$ so that $i=1,2,3$ denotes the $SU(3)^{(0)'}$ index and they are effectively three scalars transforming as the fundamental representation: $\Sigma_{i1},\Sigma_{i2},\Sigma_{i3}$.
The resulting classical action is given by
\begin{align}
    S_{\rm cl}^{\rm frac} = s\frac{8\pi^2}{g_c^2} + 
    \lmk 2\pi^2 \rho^2  \sum_{i=1}^{2}\sum_{n=1}^3 \vev{\Sigma}_{in}\rmk \times 2 = s\frac{8\pi^2}{g_c^2} + 8\pi^2 \rho^2 v_3^2\, ,
\end{align}
where $\vev{\Sigma}_{in}=v_3\delta_{in}$ and the factor of $2$ on the right-hand side of the first equality counts the contributions from both $\Sigma_{r+1}$ and $\Sigma_{r+s}$.
Hence, the instanton amplitudes of such ``fractional'' instanton backgrounds receive the exponential suppression from the Higgsing,
\begin{align}
    W_I^{\rm frac} \propto e^{-8\pi^2\rho^2 v_3^2}\, .
\end{align}
Since the instanton size that is relevant to the 5D gauge theory satisfies
\begin{align}
    R^{-1} < \rho^{-1} < \Lambda_5 \ll v_3 \, ,
\end{align}
such instanton effects can be neglected in the 5D regime.
Therefore, it is consistent that the ``fractional" instantons are included in the deconstruction setup as a UV completion of the 5D theory.

Finally, let us discuss the region $\rho \ll \Lambda_5^{-1}$, which lies outside the regime where the 5D effective description is reliable and should instead be analyzed within the full deconstruction theory.

In the range $\rho \ll \Lambda_5^{-1}$, the effective instanton action $S_{\rm eff}^I(\rho)$ for the zero-mode instanton was studied in Ref.~\cite{Poppitz:2002ac}. Their result suggests that the instanton contribution continues to grow as $\rho^{-1}$ increases. 
For the ``fractional'' instantons, Ref.~\cite{Poppitz:2002ac} argued that their contributions should be regarded as lattice artifacts and therefore discarded. However, if deconstruction is instead viewed as a particular UV completion whose infrared limit reproduces the 5D KK theory, the fractional instantons may also be interpreted as genuine UV-sensitive contributions. From this perspective, they could affect the axion quality problem. In particular, the exponential suppression present in the 5D regime is expected to disappear once $\rho$ becomes comparable to $v_3^{-1}$, so these contributions become non-negligible in the deconstructed theory.

To conclude, understanding the behavior of $S_{\rm eff}^I(\rho)$ in the region $\rho \ll \Lambda_5^{-1}$ is important for clarifying the role of instantons in the axion quality problem within deconstructed extra-dimensional axion models. We leave a more detailed analysis for a future work.

\subsection{Bulk Charged Fields}

Let us consider another source of the axion potential: $U(1)$-charged matter fields propagating in the bulk.
We already introduced a bulk scalar $Q(x,y)$ which is charged under the $U(1)$ gauge symmetry in Sec.~\ref{sec:review} and presented its deconstruction realization in Sec.~\ref{sec:deconstruction}.
As explained around Eq.~\eqref{Q_redefinition}, all the Goldstone dependence can be rotated away from the scalar potential, which means the bulk charged scalar with the minimal action~\eqref{deconst_scalar} contributes nothing to the axion potential and is harmless to the axion quality.
This is expected since the $U(1)_{\rm shift}$ is preserved in the scalar sector.

However, due to the Dirichlet boundary conditions on $A_\mu(x,y)$ and the resulting absence of the $U(1)$ gauge symmetries on the boundaries, we can write down $U(1)_{\rm shift}$-violating terms such as
\begin{align}
    S_5^{\rm bound} = &-\int\dd^4x \lmk \mu_{0}Q(x,0)+ M_0 Q(x,0)^2  + \mathrm{h.c.}\rmk \notag \\
    &- \int\dd^4x \lmk \mu_{\pi R}Q(x,\pi R)+ M_{\pi R} Q(x,\pi R)^2 +\mathrm{h.c.} \rmk \, . \label{5D_scalar_boundary}
\end{align}
Here, since the mass dimension of the field $Q$ is $\frac{3}{2}$, the mass dimensions of the coefficients are $[\mu_{0,\pi R}]=\frac{5}{2},~[M_{0,\pi R}]=1$.
Correspondingly in the deconstruction picture, we introduce
\begin{align}
	S_{D}^{\rm bound} = -\int \dd^4 x \lmk  \mu_1 Q_1 + \mu_N Q_N + \mathrm{h.c.}\rmk - \int \dd^4 x \lmk  M_1 Q_1^2 + M_N Q_N^2 + \mathrm{h.c.}\rmk\, , \label{deconst_scalar_boundary}
\end{align}
with the correspondence in the coefficients,
\begin{align}
	\mu_{1,N} \longleftrightarrow \mu_{0,\pi R}/\sqrt{a} \,, \quad M_{1,N} \longleftrightarrow M_{0,\pi R}/a\, .
\end{align}
Together with the correspondence~\eqref{scalar_correspondence}, the 5D boundary terms~\eqref{5D_scalar_boundary} are reproduced exactly.
One subtlety is that in the continuum limit, $\mu_{1,N}\sim {a}^{-1/2},~M_{1,N}\sim a^{-1}$ diverge with keeping $\mu_{0,\pi R},~M_{0,\pi R}$ finite.
After the boundary operators~\eqref{deconst_scalar_boundary} are introduced, the field redefinition~\eqref{Q_redefinition} will also change them and the axion cannot be completely removed from the scalar potential.
We perform the field redefinition and rotate all the Goldstone dependence to the $Q_N$ boundary terms, so that the full scalar potential is given by
\begin{align}
    V(Q) =  Q_i^* \lmk \mathcal{M}_Q^2\rmk_{ij} Q_j + \lmk \mu_1 Q_1 + M_1 Q_1^2 + \mu_N e^{-i\theta(x)} Q_N + M_N e^{-2i\theta(x)}Q_N^2 +\mathrm{h.c.}\rmk \, ,
\end{align}
where we replace $\sum_j \chi_j/v$ in Eq.~\eqref{Q_redefinition} by the dimensionless axion field $\theta(x)$ according to the definition~\eqref{deconst_theta}.
For simplicity, we will calculate the axion potential induced by $Q_j(x)$ when either the linear term or the quadratic term is turned on.

\subsubsection{Linear $U(1)_{\rm shift}$-violating Term}
When only the linear term is turned on, the $\theta$-dependent scalar potential can be written in a compact form as
\begin{align}
	V^{\rm lin}(\mathcal{Q};\theta) = \mathcal{Q}^\dag K \mathcal{Q} + J^\dag(\theta) \mathcal{Q} + \mathcal{Q}^\dag J(\theta) \, ,
\end{align}
where $\mathcal{Q}$ is the column vector consisting of all the $Q_j$ fields, $K\equiv \mathcal{M}_Q^2$ is just a simplified notation for the mass matrix and $J(\theta)$ is the linear source depending on the axion,
\begin{align}
\mathcal{Q} = \begin{pmatrix}
    Q_1(x) \\ Q_2(x) \\ \vdots \\ Q_{N-1}(x) \\Q_N(x)
\end{pmatrix} , 
	K = \begin{pmatrix}
M_Q^2+t & -t & 0 & \cdots & 0\\
-t & M_Q^2+2t & -t & \cdots & 0\\
0 & -t & M_Q^2+2t & \ddots & \vdots\\
\vdots & & \ddots & \ddots & -t\\
0 & 0 & \cdots & -t & M_Q^2+t
\end{pmatrix} , J=\begin{pmatrix}
		\mu_1^* \\
		0 \\
		\vdots \\
		0 \\
		\mu_N^* e^{i\theta(x)}
	\end{pmatrix} , \notag
\end{align}
with $t\equiv g^2 v^2 $.
Taking $\theta$ as a constant background field, we can find the effective potential by integrating out $Q_j(x)$ in the path integral,
\begin{align}
	V^{\rm lin}_{\rm eff}(\theta) = - \frac{1}{\mathcal{V}_4}\ln Z^{\rm lin}(\theta)\, ,
\end{align}
with the generating functional defined as
\begin{align}
	Z^{\rm lin}(\theta) &= \int D \mathcal{Q} D \mathcal{Q}^\dag e^{-S^{\rm lin}_E[Q,\theta]}\\
	 &=\int D \mathcal{Q} D \mathcal{Q}^\dag \exp\lkk \int \dd^4 x\lmk -\mathcal{Q}^\dag \mathcal{O} \mathcal{Q} - J^\dag(\theta) \mathcal{Q} - \mathcal{Q}^\dag J(\theta)\rmk \rkk \, ,
\end{align}
where $\mathcal{O}_{ij} =-\del^2\delta_{ij} + K_{ij}$ is the kernel and we neglect the gauge field couplings and the derivative terms of $\chi_j$ induced by the field redefinition since they do not contribute to the effective potential of $\theta$. Performing the Gaussian integral, we obtain
\begin{align}
	Z^{\rm lin}(\theta) = \mathcal{N}_\pi \lmk\mathrm{Det}\mathcal{O}\rmk^{-1} \exp\lmk \int \dd^4x J^\dag(\theta) K^{-1} J(\theta)\rmk \, .
\end{align}
Here, $\mathcal{N}_\pi=\Pi_A \pi$ denotes the formal product counting the dimension of the eigensystem of $\mathcal{O}$, $\mathrm{Det}$ is the functional determinant and we use the fact that $J(\theta)$ has no spacetime dependence to reduce the full kernel to the mass matrix in the exponent.
Therefore, the effective action can be written as
\begin{align}
	V^{\rm lin}_{\rm eff}(\theta) &= V_0 - J^\dag (\theta)_i \lmk K^{-1}\rmk_{ij} J(\theta)_j \notag \\[1ex]
	& = V_0 - |\mu_1|^2 \lmk K^{-1}\rmk_{11} - |\mu_N|^2 \lmk K^{-1}\rmk_{NN} - \mu_1\mu_N^* e^{i\theta} \lmk K^{-1}\rmk_{1N} - \mu_1^*\mu_N e^{-i\theta} \lmk K^{-1}\rmk_{N1} \notag\\[1ex]
	& = V_0' - 2\Re\lkk \mu_1\mu_N^* e^{i\theta} \lmk K^{-1}\rmk_{1N} \rkk \label{Veff_linear_0} \, ,
\end{align} 
where $V_0 = -\frac{1}{\mathcal{V}_4}\ln \lmk \mathcal{N}_\pi  \lmk\det\mathcal{O}\rmk^{-1}\rmk$ and $V_0' = V_0 - |\mu_1|^2 \lmk K^{-1}\rmk_{11} - |\mu_N|^2 \lmk K^{-1}\rmk_{NN}$ are $\theta$-independent and neglected below. 
In the last equality, we have also used the symmetric property of the matrix $K$ so that $\lmk K^{-1}\rmk_{1N}= \lmk K^{-1}\rmk_{N1}$.
The calculation of the inverse of the matrix $K$ is straightforward since we already solved for its eigensystems~\eqref{eigenmass_scalar}, \eqref{eigenstate_scalar}, which can be written in a more formal way as
\begin{align}
	\lambda_k = M_Q^2 + 4t\sin^2\lmk\frac{k\pi}{2N}\rmk, \quad u_j^{(k)} = \sqrt{\frac{c_k}{N}}\cos\lmk \frac{(j-\half)k\pi}{N}\rmk \, ,
\end{align}
with $k=0,\cdots, N-1$. Then, the inverse matrix can be simply calculated as
\begin{align}
	\lmk K^{-1}\rmk_{ij} = \sum_{k=0}^{N-1}\frac{u_i^{(k)}u_j^{(k)}}{\lambda_k}\, ,
\end{align}
and therefore,
\begin{align}
	\lmk K^{-1}\rmk_{1N} = \sum_{k=0}^{N-1} \frac{u_1^{(k)}u_N^{(k)}}{\lambda_k} = \frac{\sinh \alpha}{M_Q^2 \sinh\lmk N\alpha \rmk}\, ,
\end{align}
where $\alpha$ is defined as
\begin{align}
	\cosh\alpha = 1+\frac{M_Q^2}{2t} = 1+\frac{M_Q^2}{2g^2 v^2} \, .
\end{align}
Finally, we obtain the effective potential,
\begin{align}
	V^{\rm lin}_{\rm eff}(\theta) = - \frac{2\abs{\mu_1\mu_N} \sinh \alpha}{M_Q^2 \sinh\lmk N\alpha \rmk}\cos\lmk\theta+\delta_\mu\rmk\, ,\label{Veff_linear}
\end{align}
with $\delta_\mu = \arg(\mu_1)-\arg(\mu_N)$. 
Since in the above calculation, no momentum integral has been performed to derive the effective potential in the axion background, 
it turns out to be a classical effect.
In fact, after the introduction of the linear terms in $Q_1,~Q_N$, all the scalar fields develop non-zero VEVs.
$\vev{Q_N}\neq 0$ together with the $\theta$-dependent linear term gives the axion potential in a purely classical way and the result is the same as Eq.~\eqref{Veff_linear}.
The two approaches are equivalent while the one performing path integral can be easily generalized to the case of the quadratic term.

We are interested in the near-continuum limit region where $gv\gg M_Q$ and $\alpha\simeq 0$. 
We can Taylor expand $\cosh\alpha\simeq 1+\frac{\alpha^2}{2}$ and find that $\alpha\simeq \frac{M_Q}{gv}$. Then the effective potential is reduced to
\begin{align}
	V^{\rm lin}_{\rm eff}(\theta)\simeq -\frac{2\abs{\frac{\mu_1\mu_N}{gv}} M_Q}{M_Q^2 \sinh\lmk N\alpha \rmk}\cos\lmk\theta+\delta_\mu\rmk\simeq -\frac{4\abs{\mu_1\mu_N}}{gv M_Q}e^{-M_Q\pi R}\cos\lmk\theta+\delta_\mu\rmk\, , \label{Veff_linear_largev}
\end{align}
where we use the fact that $N/(gv)$ is fixed to be $\pi R$ in the deconstruction setup. In the continuum limit, Eq.~\eqref{Veff_linear_largev} corresponds to the 5D potential,
\begin{align}
	\lim_{N\to \infty} V_{\rm eff}(\theta) \, \longleftrightarrow \, V_5(\theta)=-\frac{4\abs{\mu_0\mu_{\pi R}}}{M_Q}e^{-M_Q\pi R}\cos\lmk\theta+\delta_\mu\rmk \, .
\end{align}
The result appears somewhat surprising because the origin of the axion potential is the $U(1)$-violating operators localized on the two endpoints (branes) but the resulting potential is still exponentially suppressed, which is typical of non-local effects.
This becomes clear upon examining the calculations more closely.
First, $U(1)$-violating operators must be present at both endpoints otherwise the Goldstone dependence can be shifted entirely to the $U(1)$-preserving endpoint and therefore drops out of the scalar potential.
Second, the effective potential is proportional to $\lmk K^{-1}\rmk_{1N}$ in Eq.~\eqref{Veff_linear_0}, which is nothing but the propagator connecting the two ends of the lattice.
Therefore, Eqs.~\eqref{Veff_linear}, \eqref{Veff_linear_largev} should be understood as non-local effects: they arise from the propagation across the full lattice that communicates the $U(1)$-violation localized at one endpoint to the other.

\subsubsection{Quadratic $U(1)_{\rm shift}$-violating Term}
The scalar potential with the boundary quadratic terms can be written as
\begin{align}
	V^{\rm quad}(\mathcal{Q};\theta) = \mathcal{Q}^\dag K \mathcal{Q} + \half \mathcal{Q}^T B\mathcal{Q} + \half \mathcal{Q}^\dag B^\dag \mathcal{Q}^* \, ,
\end{align}
where the $N\times N$ matrix $B$ is defined as
\begin{align}
    B(\theta) = \mathrm{diag} (2M_1,0,\cdots,0,2M_N e^{-i2\theta(x)})\, .
\end{align}
We introduce the doubled field,
\begin{align}
    \Xi = \begin{pmatrix}
        \mathcal{Q} \\ \mathcal{Q}^*
    \end{pmatrix}\, ,\quad 
    \Xi^\dag = \begin{pmatrix}
        \mathcal{Q}^\dag & \mathcal{Q}^T
    \end{pmatrix}\, ,
\end{align}
and the Euclidean action can be written in the Gaussian form,
\begin{align}
    S_E^{\rm quad} = \frac{1}{2} \int\dd^4 x ~\Xi^\dag(x) \mathbb{O}(\theta) \Xi(x)\, ,
\end{align}
where the operator $\mathbb{O}$ is defined as
\begin{align}
    \mathbb{O} = \begin{pmatrix}
        \mathcal{O} & B^\dag(\theta) \\
        B(\theta) & \mathcal{O}^T
    \end{pmatrix} \, .
\end{align}
As the inverse propagator is $\theta$-dependent, let us move to the momentum space for the calculation of the effective potential,
\begin{align}
    \mathcal{Q}(x) = \int\frac{\dd^4 p}{(2\pi)^2} e^{ipx}\widetilde{\mathcal{Q}}(p)\, ,\quad \mathcal{Q}^\dag (x) = \int\frac{\dd^4 p}{(2\pi)^2} e^{-ipx}\widetilde{\mathcal{Q}}^\dag(p) \, .
\end{align}
Now the Euclidean action is written as
\begin{align}
    S_E^{\rm quad} = \half \int\frac{\dd^4 p}{(2\pi)^4} \widetilde{\Xi}^\dag (-p) \widetilde{\mathbb{O}}(p,\theta)\widetilde{\Xi}(p) \, ,
\end{align}
where
\begin{align}
    \widetilde{\Xi}(p) = \begin{pmatrix}
        \widetilde{\mathcal{Q}}(p) \\ \widetilde{\mathcal{Q}}^*(p)
    \end{pmatrix} \, , \quad 
  \widetilde{\mathbb{O}}(p,\theta) = \begin{pmatrix}
        \widetilde{\mathcal{O}}(p,\theta) & B^\dag(\theta) \\
        B(\theta) & \widetilde{\mathcal{O}}^T(p,\theta)
    \end{pmatrix} = 
    \begin{pmatrix}
        p^2 \mathbf{1}_N+K & B^\dag (\theta) \\
        B(\theta) & p^2 \mathbf{1}_N+K
    \end{pmatrix} \, .
\end{align}
Here we already took $\theta$ a constant background field and the generating functional can be calculated as
\begin{align}
    Z^{\rm quad}(\theta) =\int D \widetilde{\mathcal{Q}} D \widetilde{\mathcal{Q}}^\dag e^{ -\half \int \frac{\dd^4 p }{(2\pi)^4} ~\widetilde{\Xi}^\dag(-p) \widetilde{\mathbb{O}}(p,\theta) \widetilde{\Xi}(p) } \propto \lkk \mathrm{Det} \widetilde{\mathbb{O}}(p,\theta)\rkk^{-\half} \, ,
\end{align}
where $\mathrm{Det}$ is the functional determinant counting different values of the momentum and we neglect the prefactor which becomes a constant in the effective potential, as in the case of the linear term.
Then, the effective potential is given by
\begin{align}
    V_{\rm eff}^{\rm quad}(\theta) = \frac{1}{2\mathcal{V}_4}  \ln \lmk \mathrm{Det} \lkk \widetilde{\mathbb{O}}(p,\theta)\rkk \rmk +\mathrm{const} 
    = \frac{1}{2\mathcal{V}_4}  \Tr \lmk \ln \lkk \widetilde{\mathbb{O}}(p,\theta)\rkk \rmk +\mathrm{const}\, ,
\end{align}
where $\Tr$ is the functional trace which traces over the momentum eigenstates and the internal matrix index,
\begin{align}
    \Tr \lmk \ln \lkk \widetilde{\mathbb{O}}(p,\theta)\rkk \rmk 
    =\sum_p \tr \lmk\ln\lkk \widetilde{\mathbb{O}}(p,\theta)\rkk\rmk 
    = \mathcal{V}_4\int\frac{\dd^4 p}{(2\pi)^4}  \ln\lmk \det \lkk\widetilde{\mathbb{O}}(p,\theta)\rkk\rmk\, .
\end{align}
Here, $\tr$ denotes the ordinary trace over the $2N\times 2N$ matrix and we turn it back to the determinant in the last equality.
The determinant of $\widetilde{\mathbb{O}}$ can be expanded in terms of $\widetilde{\mathcal{O}}$ and $B(\theta)$ using the Schur complement formula,
\begin{align}
    \det \widetilde{\mathbb{O}} &= \det (\widetilde{\mathcal{O}}) \det\lmk \widetilde{\mathcal{O}}^T - B(\theta) \widetilde{\mathcal{O}}^{-1} B(\theta)^\dag\rmk \notag \\
    &= \det (\widetilde{\mathcal{O}})^2 \det\lmk \mathbf{1}_N - \widetilde{\mathcal{O}}^{-1}B(\theta) \widetilde{\mathcal{O}}^{-1} B(\theta)^\dag\rmk \, ,
\end{align}
where we use the fact that $\widetilde{\mathcal{O}}$ is symmetric.
Since the prefactor does not depend on $\theta$, it just contributes to a constant in the effective potential.
Now neglecting all the irrelevant constants, the effective potential of the axion is given by
\begin{align}
    V_{\rm eff}^{\rm quad}(\theta) = \half \int\frac{\dd^4 p}{(2\pi)^4} \ln\lkk \det\lmk \mathbf{1}_N - \widetilde{\mathcal{O}}^{-1}B(\theta) \widetilde{\mathcal{O}}^{-1} B(\theta)^\dag\rmk \rkk \equiv \half \int\frac{\dd^4 p}{(2\pi)^4} \ln D(p,\theta)\, .
\end{align}
Since the matrix $B(\theta)$ only has non-zero elements at two endpoints, the determinant can be much simplified to that of a $2\times 2$ matrix.
Define
\begin{align}
f(p)\equiv \lmk \widetilde{\mathcal{O}}^{-1}\rmk_{11}(p)=\lmk \widetilde{\mathcal{O}}^{-1}\rmk_{NN}(p) \, ,
\quad
h(p)\equiv \lmk \widetilde{\mathcal{O}}^{-1}\rmk_{1N}(p)=\lmk \widetilde{\mathcal{O}}^{-1}\rmk_{N1}(p) \, ,
\end{align}
and the $2\times 2$ matrices
\begin{align}
F(p)\equiv
\begin{pmatrix}
f(p) & h(p)\\
h(p) & f(p)
\end{pmatrix},  \quad  B_e(\theta)\equiv \begin{pmatrix}
2M_1 & 0\\
0 & 2M_N e^{-2i\theta}
\end{pmatrix}.
\end{align}
Then we find
\begin{align}
    D(p,\theta) =& \det\lmk \mathbf{1}_2-F(p) B_e F(p) B_e^\dag\rmk \notag \\[1ex]
    =& 1 -4f(p)^2\lmk |M_1|^2+|M_N|^2 \rmk -8h(p)^2\Re\lkk M_1 M_N^* e^{2i\theta}\rkk  \notag \\
    &+16|M_1M_N|^2 \lmk f(p)^2-h(p)^2\rmk^2  \, .
\end{align}
Notice that the axion dependence is again controlled by the non-local piece $h(p)$.
The evaluation of $f(p), h(p)$ is quite similar to what we have done with $K^{-1}$ since $\widetilde{\mathcal{O}}^{-1}$ is the same as $K^{-1}$ after replacing $M_Q^2$ by $p^2+M_Q^2$.
Therefore, we obtain
\begin{align}
f(p) =\frac{\sinh\beta_p \cosh\lmk (N-1)\beta_p\rmk}{(M_Q^2+p^2)\sinh(N\beta_p)} \, , \quad
h(p) =\frac{\sinh\beta_p}{(M_Q^2+p^2)\sinh(N\beta_p)}\, ,
\end{align}
with $\beta_p$ defined by
\begin{align}
    \cosh\beta_p = 1 +\frac{p^2+M_Q^2}{2t} \, .
\end{align}
Since the determinant $D(p,\theta)$ is still in a complicated form, we need to take some limit to have an analytical result.
As before we take the continuum limit, $M_Q^2\ll t$.
However, now the value of $\beta_p$ also depends on $p$ and if the momentum is large, the approximation we took before is no longer valid.
For later convenience, we extract a constant term $V_{\rm eff}^{\rm quad}(0)$ from the effective potential and then consider
\begin{align}
     V_{\rm eff}^{\rm quad}(\theta) = \half \int\frac{\dd^4 p}{(2\pi)^4} \ln \frac{D(p,\theta)}{D(p,0)} \, .
\end{align}
As mentioned, the axion dependence of the determinant is proportional to $h(p)^2$,
\begin{align}
    D(p,\theta)-D(p,0) = -8 h(p)^2 \lmk \Re\lkk M_1M_N^* e^{2i\theta}\rkk - \Re\lkk M_1M_N^* \rkk\rmk\, .
\end{align}
As long as $N\beta_p\gg 1$, which is natural in the continuous limit, we have $h(p)\ll f(p)=\cosh\lmk(N-1)\beta_p\rmk h(p)$ 
and $|D(p,\theta)-D(p,0)|\ll |D(p,0)|$.
In this case, the effective potential can be approximated as
\begin{align}
     V_{\rm eff}^{\rm quad}(\theta) &= \half \int\frac{\dd^4 p}{(2\pi)^4} \ln \lmk 1 + \frac{D(p,\theta)-D(p,0)}{D(p,0)} \rmk \notag \\[1ex]
     &\simeq  -4 \lmk \Re\lkk M_1M_N^* e^{2i\theta}\rkk - \Re\lkk M_1M_N^* \rkk\rmk\int\frac{\dd^4 p}{(2\pi)^4} \frac{ h(p)^2 }{D_{\rm loc}(p)} \, , \label{Veff_quad_no_approx}
\end{align}
where in the denominator we further approximate $D(p,0)$ as $D_{\rm loc}(p)$ by subtracting all the terms proportional to $h(p)^2$,
\begin{align}
    D_{\rm loc}(p)\equiv 1 -4f(p)^2\lmk |M_1|^2+|M_N|^2 \rmk +16|M_1M_N|^2 f(p)^4 \, .
\end{align}
We will take such approximation below.

First we consider the large-momentum regime $p^2\gg t$ to check whether the integral converges or not.
In this case, $\beta_p$ is also large and we approximately find
\begin{align}
    \cosh\beta_p\simeq \half e^{\beta_p}\simeq \frac{p^2}{2t} \, ,\quad \sinh\beta_p\simeq \half e^{\beta_p}\simeq \frac{p^2}{2t} \, .
\end{align}
Then $f(p), h(p)$ can be estimated as
\begin{align}
    f(p)\simeq \frac{1}{p^2} \, , \quad  h(p) \simeq \frac{(2t)^{N-1}}{p^{2N}} \, .
\end{align}
Since $M_{1,N}$ scales as $a^{-1}=\sqrt{t}$, they are also much smaller than $p^2$ and thus $D_{\rm loc}(p)\simeq 1$.
Therefore, we obtain
\begin{align}
    V_{\rm eff}^{\rm quad}(\theta) \supset -4 \lmk \Re\lkk M_1M_N^* e^{2i\theta}\rkk - \Re\lkk M_1M_N^* \rkk\rmk \int_{p^2\gg t} \frac{\dd^4 p}{(2\pi)^4} \frac{(2t)^{2N-2}}{p^{4N}} \, .
\end{align}
Obviously, the integral is significantly suppressed at large momentum when $N$ is large.

Now we turn to the small-momentum regime $p^2\ll t$, where $\cosh\beta_p\simeq 1+\half\beta_p^2$ and we have
\begin{align}
    \beta_p \simeq \sqrt{\frac{p^2+M_Q^2}{t}} \, .
\end{align}
In addition, we consider the large mass regime where $M_Q\pi R\gg 1$ so that 
\begin{align}
    N\beta_p = \frac{N}{gv} \sqrt{p^2 +M_Q^2}> M_Q\pi R \gg 1 \, .
\end{align}
Then we get
\begin{align}
    f(p)\simeq \frac{\beta_p e^{(N-1)\beta_p}}{(M_Q^2+p^2)e^{N\beta_p}} = \frac{e^{-\beta_p}}{\sqrt{t(M_Q^2+p^2)}} \, , \\[1ex]
    h(p) \simeq \frac{2\beta_p}{M_Q^2+p^2} e^{-N\beta_p} =\frac{2e^{-\pi R\sqrt{p^2+M_Q^2}}}{\sqrt{t(M_Q^2+p^2)}} \, .
\end{align}
In order to find an analytic form, we have to work in the region of $p\ll M_Q$, where the numerator of the integrand is proportional to
\begin{align}
    h(p)^2\simeq 4\frac{e^{-2\pi R \lmk M_Q +\frac{p^2}{2M_Q}\rmk}}{tM_Q^2} \lmk 1- \frac{p^2}{M_Q^2}\rmk = 4\frac{e^{-2\pi RM_Q}}{tM_Q^2}e^{-\frac{\pi R}{M_Q}p^2} + O\lmk \frac{p^2}{M_Q^2} \rmk \, ,
\end{align}
and the denominator is approximated as $D_{\rm loc}(p)\simeq D_{\rm loc}(0)$.
Finally, we obtain a Gaussian integral over $p$,
\begin{align}
    V_{\rm eff}^{\rm quad}(\theta) &\supset -16 \frac{\lmk \Re\lkk M_1M_N^* e^{2i\theta}\rkk - \Re\lkk M_1M_N^* \rkk\rmk}{D_{\rm loc}(0)} \frac{e^{-2\pi RM_Q}}{tM_Q^2} \int_{p\ll M_Q} \frac{p^3 \dd p} {8\pi^2} e^{-\frac{\pi R}{M_Q}p^2} \notag \\[1ex]
    &=-\frac{2\lmk \Re\lkk M_1M_N^* e^{2i\theta}\rkk - \Re\lkk M_1M_N^* \rkk\rmk e^{-2\pi R M_Q}}{\pi^2 t M_Q^2 D_{\rm loc}(0)} \frac{M_Q^2}{2 (\pi R)^2} \notag \\[1ex]
    &=-\frac{2|M_0 M_{\pi R}^*|}{\pi^4 R^2\lmk1-\frac{4|M_0|^2}{M_Q^2}\rmk \lmk1-\frac{4|M_{\pi R}|^2}{M_Q^2}\rmk}e^{-2\pi R M_Q} \cos\lmk 2\theta +\delta_M\rmk\, , \label{Veff_quad}
\end{align}
where $D_{\rm loc}(0)$ is estimated as
\begin{align}
    D_{\rm loc}(0) &= 1 -4f(0)^2\lmk |M_1|^2+|M_N|^2 \rmk +16|M_1M_N|^2 f(0)^4 \notag \\[1ex]
    &= (1-4f(0)^2|M_1|^2)(1-4f(0)^2|M_N|^2) \notag \\[1ex]
    &\simeq \lmk1-\frac{4|M_0|^2}{M_Q^2}\rmk \lmk1-\frac{4|M_{\pi R}|^2}{M_Q^2}\rmk \, ,
\end{align}
with $f(0)\simeq \frac{1}{\sqrt{t}M_Q}$.
The exponential factor $e^{-2\pi RM_Q}$ originates from the non-local piece $h(p)\equiv \lmk\tilde{\mathcal{O}}^{-1}\rmk_{1N}(p)$, which implies that such effective axion potential is due to the non-local effects, again.

So far, we have focused on the cases where the mass of the bulk scalar $M_Q$ is much smaller than the inverse of the lattice interval $\sqrt{t}=a^{-1}$, which is reasonable for a scalar considered in the 5D theory, $M_Q < \Lambda_5$.
It turns out that both axion potentials~\eqref{Veff_linear_largev}, \eqref{Veff_quad} calculated in the deconstruction setup are consistent with the results derived in the 5D picture~\cite{Choi:2026kxu}.
The results show that with the boundary operators unsuppressed, $\mu_{0,\pi R}\simeq (\pi R)^{-5/2},~M_{0,\pi R}\simeq (\pi R)^{-1}$, the bulk mass is required to be larger than the compactification scale so that the exponential factors are sufficiently small and the axion quality is ensured.

However, from the perspective of the deconstructed theory, the scalar with a larger mass should also be considered.
In this case, $M_Q$ can be comparable to $\sqrt{t}$ and the approximation for small $\alpha,\beta_p$ fails.
In addition, the coefficients of the boundary operators $\mu_{1,N}, M_{1,N}$ can also be as large as $\sqrt{t}^3,\sqrt{t}^2$.
Nevertheless, by inspecting the expression before the approximation~\eqref{Veff_linear}, \eqref{Veff_quad_no_approx}, we find that as $M_Q$ increases, both of the effective potentials decrease monotonically, with the ratios between the coefficients of the boundary operators and the bulk mass $\frac{|\mu_{1,N}|}{M_Q^3},~\frac{|M_{1,N}|}{M_Q^2}$ fixed.
In conclusion, the heavy matter field that might appear only in the deconstruction setup is harmless to the axion quality compared with the lighter field in the 5D effective theory.
\section{Conclusions and discussions
\label{sec:Discussion}}

In the present work, we have constructed a four-dimensional deconstruction of the extra-dimensional axion based on a moose of $U(1)\times SU(3)$ gauge theory. The setup reproduces the essential features of the 5D orbifold construction in which the axion arises as the Wilson line of a bulk $U(1)$ gauge field and couples to QCD through a Chern-Simons term. In the 4D deconstructed theory, the axion emerges as a collective pNGB associated with the phases of the link fields, providing a fully four-dimensional and renormalizable framework for the extra-dimensional axion.
We have demonstrated that the correspondence between the five-dimensional and deconstructed descriptions extends beyond the perturbative KK spectrum. In particular, the axion-gluon coupling is reproduced by a gauged Wess-Zumino-Witten term, providing a precise 4D counterpart of the 5D Chern-Simons interaction. 
We have also analyzed the non-perturbative effects in the deconstructed theory. 
For instantons whose size lies within the regime admitting a 5D interpretation, the dominant contribution is the ordinary instanton associated with the diagonal gauge group. 
The site-localized ``fractional'' instantons, although present as legitimate configurations in the discretized theory, give exponentially suppressed contributions in the continuum limit and therefore do not spoil the recovery of the higher-dimensional result.
We further discussed that smaller instantons, whose inverse size lies above the 5D cutoff, probe the intrinsically deconstructed regime, where the total instanton-induced axion potential may continue to grow and reach a maximum value larger than the estimate based only on the 5D effective description.
Furthermore, we have studied axion potentials generated by bulk matter fields and boundary-localized symmetry-breaking operators. The resulting potentials exhibit the characteristic nonlocal suppression associated with the propagation across the extra dimension. From the deconstruction viewpoint, this suppression originates from the collective structure of the multi-site theory and the requirement of multiple link-field insertions connecting distant sites. This provides a transparent four-dimensional understanding of the geometrical protection mechanism underlying the high quality of the extra-dimensional axion.
Our construction demonstrates that the essential features of the extra-dimensional axion can survive within a four-dimensional UV completion. The deconstruction framework offers a useful setting in which non-perturbative effects and possible sources of PQ-symmetry breaking can be analyzed quantitatively with better theoretical control.

The present setup can also be extended to warped backgrounds, such as the Randall-Sundrum geometry
\cite{Randall:1999ee}. In the deconstruction picture, this would correspond to introducing site-dependent VEVs of link fields
\cite{Cheng:2001nh,Abe:2002rj,Falkowski:2002cm,Randall:2002qr,deBlas:2006fz,Burdman:2012sb,Nakai:2014iea}. Such a construction may modify the localization of the axion mode and the structure of non-perturbative effects, potentially providing a different perspective on the axion quality problem.

Since the five-dimensional theory should be regarded only as an effective description below a UV cutoff scale, the post-inflationary axion scenario—defined as the regime in which the effective axion decay constant is not a fundamental input but is determined by the dynamics of underlying radial (modulus) fields—necessarily involves physics beyond the regime of validity of the 5D effective theory.
For this reason, extra-dimensional axion models are often discussed in the pre-inflationary scenario. In contrast, the deconstructed theory provides a fully four-dimensional and renormalizable UV completion in which the post-inflationary dynamics can, in principle, be studied explicitly.

In explicit UV completions of extra-dimensional axion models, cosmic strings associated with moduli stabilization have recently been investigated in Refs.~\cite{March-Russell:2021zfq,Benabou:2023npn,Cline:2024vbd}.
In such constructions, the axion direction disappears in the string core through its interplay with moduli dynamics, leading to a decompactification of the extra dimension. As a result, the string tension is not controlled by the low-energy axion decay constant and can be parametrically larger. Interestingly, cosmic strings in the deconstructed picture exhibit qualitatively similar features. 
In the deconstructed theory, the vacuum expectation values of the link fields set the lattice spacing of each link, and the total length of the moose determines the size of the emergent extra dimension.
At the core of a cosmic string, the link fields vanish, corresponding to a deconstruction breakdown or decompactification limit. Furthermore, the string tension is controlled by the scale associated with the lattice spacing rather than by the effective axion decay constant determined by the total length of the moose.

These observations suggest that the deconstructed framework may provide a useful setting for studying post-inflationary axion cosmology in UV-complete realizations of extra-dimensional axions. One may further consider finite-site models and investigate explicit string solutions, the interaction energy between strings, and the associated domain wall number, i.e. how the axion winding is distributed around the string configuration. Such studies may open a new direction toward deconstruction-inspired post-inflationary axion models that simultaneously address the axion quality problem and cosmological issues.


\section*{Acknowledgments}

Y.N. is supported by Natural Science Foundation of Shanghai. M.S. is supported by the MUR projects 2017L5W2PT.
M.S. also acknowledges the European Union - NextGenerationEU, in the framework of the PRIN Project “Charting unexplored avenues in Dark Matter” (20224JR28W).
M.S. thanks Arthur Platschorre for useful discussions on cosmic string formation in 5D.


\appendix
\section{Derivation of the 5D Spectrum\label{sec:appendix_5D_spectrum}}
Here we explicitly show the derivation of the 5D EoMs, general boundary conditions and the resulting KK spectrum for the bulk gauge fields $A_M(x, y), G_M(x,y)$ and the matter fields $\Psi(x,y), Q(x,y)$.

\subsection{Gauge Sector}

Performing the variation upon the 5D action of the gauge fields~\eqref{5D_gauge} gives
\begin{align}
\delta S_{5}^{\rm gauge}
&=\frac{1}{g_5^2}\left\{\int d^5x\lkk\lmk\del_M \mathcal{F}^{M\nu}+\xi_1^{-1}\del^\nu \del^\mu A_\mu-\del^\nu\partial_y A_5\rmk \delta A_\nu\right. \notag \right.\\
&\left. + \lmk\partial_\mu \mathcal{F}^{\mu 5}-\partial_y \del^\mu A_\mu - \xi_1 \del_y \del^y A_5\rmk\delta A_5 \rkk \notag \\
& \left. + \int d^4x\lkk\lmk\partial_y A^\mu+\xi_{1,\rm b}^{-1}\partial^\mu\partial^\nu A_\nu\rmk\delta A_\mu
+\lmk \xi_1\partial_y A_5\mp \xi_{1,\rm b}A_5\rmk \delta A_5\rkk^{\pi R}_0 \right\} \notag \\
& + \frac{1}{g_{5,c}^2}\left\{\int d^5x\lkk\lmk \lmk D_M \mathcal{G}^{M\nu} \rmk^a +\xi_c^{-1}\del^\nu \del^\mu G^a_\mu-\del^\nu\partial_y G_5^a \rmk \delta G^a_\nu \right.  \right.\notag\\
&\left. + \lmk \lmk D_\mu \mathcal{G}^{\mu 5} \rmk^a - \partial_y \del^\mu G^a_\mu - \xi_c \del_y \del^y G_5^a \rmk\delta G_5^a \rkk \notag \\
& \left. + \int d^4x\lkk\lmk\partial_y G^{a\mu}+\xi_{c,\rm b}^{-1}\partial^\mu\partial^\nu G^a_\nu\rmk\delta G^a_\mu
+\lmk \xi_c\partial_y G_5^a \mp \xi_{c,\rm b}G^a_5\rmk \delta G^a_5\rkk^{\pi R}_0 \right\}  \, .
\label{gauge_variation}
\end{align}
The resulting bulk EoMs are
\begin{align}
\partial_M \mathcal{F}^{M\nu}+\frac{1}{\xi_1}\partial^\nu\partial^\mu A_\mu-\partial^\nu\partial_5 A_5 =0\, ,&\quad 
\partial_\mu \mathcal{F}^{\mu 5}+\partial_5\partial^\mu A_\mu-\xi_1\partial_5^2A_5 =0\, , \\
(D_M\mathcal{G}^{M\nu})^a+\frac{1}{\xi_c}\partial^\nu\partial^\mu G_\mu^a -\partial^\nu\partial_5G_5^a =0 \, , &\quad (D_\mu \mathcal{G}^{\mu 5})^a +\partial_5\partial^\mu G_\mu^a -\xi_c\partial_5^2G_5^a =0 \, .
\end{align}
The vanishing of the boundary variations requires
\begin{align}
    \lmk\partial_y A^\mu+\xi_{1,\rm b}^{-1}\partial^\mu\partial^\nu A_\nu\rmk\delta A_\mu\big|_{0,\pi R} =0 \, ,\quad \lmk \xi_1\partial_y A_5\mp \xi_{1,\rm b}A_5\rmk \delta A_5\big|_{0,\pi R} = 0 \, ,
    \label{boundary_variation_u1}\\[1ex]
    \lmk\partial_y G^{a\mu}+\xi_{c,\rm b}^{-1}\partial^\mu\partial^\nu G^a_\nu\rmk\delta G^a_\mu\big|_{0,\pi R} =0 \, ,\quad \lmk \xi_c\partial_y G^a_5\mp \xi_{c,\rm b}G^a_5\rmk \delta G^a_5\big|_{0,\pi R} = 0 \, ,\label{boundary_variation_su3}
\end{align}
where we have two choices for each condition: requiring either the variation of the field or the coefficient to vanish.
We take the Feynman gauge for the bulk gauge parameters, namely,
\begin{align}
    \xi_1 = \xi_c =1 \, ,
\end{align}
so that the bulk EoMs are reduced to 
\begin{align}
    (\Box_4-\partial_y^2)A_\mu=0\, ,\quad (\Box_4-\partial_y^2)A_5=0\, , \label{5D_EoM_u1} \\[1ex]
    (\Box_4-\partial_y^2)G^a_\mu=0\, ,\quad (\Box_4-\partial_y^2)G^a_5=0\, ,\label{5D_EoM_su3}
\end{align}
where $\Box_4\equiv \del_\mu\del^\mu$ is the 4D d’Alembertian and we linearized the EoMs for the $SU(3)$ gauge field.
On the boundaries, we let the gauge parameters be
\begin{align}
     \xi_{1,\rm b}=0\, ,\quad \xi_{c,\rm b}\to\infty\, ,
\end{align}
and impose different types of boundary conditions on the $A_M$ and $G_M$ as in Eqs.~\eqref{5D_BC_u1}, \eqref{5D_BC_su3} in order that the boundary variations~\eqref{boundary_variation_u1},~\eqref{boundary_variation_su3} vanish.

Now we can solve the wavefunctions of the KK modes by substituting the KK expansion~\eqref{KK_expansion_u1}, \eqref{KK_expansion_su3} into the bulk EoMs~\eqref{5D_EoM_u1}, \eqref{5D_EoM_su3} and the boundary conditions~\eqref{5D_BC_u1}, \eqref{5D_BC_su3}, and we find
\begin{align}
    \del_y^2 {f_n^V}(y) = m_{V,n}^2 f_n^V(y) \, ,\quad  f_n^A(y=0,\pi R)=0\, ,\quad \del_y{f_n^G}(y=0,\pi R) = 0\, ,\\[1ex]
    \del_y^2 {g_n^V}(y) = m_{V,n}^2 g_n^V(y)\, ,\quad  \del_y {g_n^A}(y=0,\pi R)=0\, ,\quad g_n^G(y=0,\pi R) = 0\, ,
\end{align}
where $V=\{A, G\}$ denotes the gauge fields. 
To make the KK modes $V_{\mu,k}$ (vectors) and $V_{5,k}$ (scalars) have mass dimension 1 as in the ordinary 4D theory, the KK profiles should be dimensionless.
Hence, we consider the normalization,
\begin{align}
   \frac{1}{\pi R}\int_0^{\pi R} \dd y\,  f_m^V(y) f_n^V(y) = \delta_{mn}\, , \quad \frac{1}{\pi R}\int_0^{\pi R} \dd y\,  g_m^V(y) g_n^V(y) = \delta_{mn} \, . \label{gauge_field_KKnorm}
\end{align}
We can then solve for the KK profiles. The result is given by
\begin{align}
	f_n^A=g_n^G &= \sqrt{2} \sin\frac{ny}{R}\, , \quad \text{for} ~ n\geq 0\, ,\\[1ex]
	g_0^A=f_0^G=1 \, &,~g_n^A=f_n^G = \sqrt{2} \cos\frac{ny}{R}\, , \quad \text{for} ~ n\geq1\, ,
\end{align}
with the KK masses solved as $m_{V,n}= \frac{n}{R}$.

\subsection{Matter Sector}
For the fermion $\Psi(x,y)$, the bulk EoMs and the general boundary conditions are derived by varying the 5D action~\eqref{5D_fermion},
\begin{align}
\delta S_5^f&=\int d^4x\int_0^{\pi R} dy\,\lkk\delta\eta^\dag\left(i\bar\sigma^\mu D_\mu\eta+D_5\psi^\dag-M_\Psi\psi^\dag\right) \right. \notag\\
&\qquad\left. +\delta\psi^\dag\left(i\bar\sigma^\mu D_\mu^*\psi-D_5\eta^\dag-M_\Psi\eta^\dag\right)+\mathrm{h.c.}\rkk \\
&+\int d^4x\,\lkk-\psi\,\delta\eta+\eta^\dag\delta\psi^\dag+\mathrm{h.c.}\rkk_0^{\pi R} .
\end{align}
We can read off the bulk EoMs,
\begin{align}
    i\bar\sigma^\mu D_\mu\eta+\left(D_5- M_\Psi\right)\psi^\dag=0 \, ,\qquad i\bar\sigma^\mu D_\mu^*\psi-\left(D_5+M_\Psi\right)\eta^\dag=0 \, ,
\end{align}
as well as the general boundary conditions,
\begin{align}
    \psi\delta\eta|_{0,\pi R} = \eta\delta\psi|_{0,\pi R} = 0 \, .
\end{align}
To make the boundary variations vanish, we only need to impose the Dirichlet boundary conditions on $\psi$,
\begin{align}
   \delta\psi|_{0,\pi R} = 0\, , \qquad \psi|_{0,\pi R}=0\, ,
\end{align}
as in Eq.~\eqref{fermion_BC}.
After the KK expansion~\eqref{KK_expansion_fermion}, we obtain the associated equations for the profiles $f_n^\eta, f_n^\psi$,
\begin{align}
    \left(\partial_y+M_\Psi\right)f_n^\eta(y)=m_{f,n} f_n^\psi(y) \, , \label{eta_KK_EoM}\\[1ex]
 \left(-\partial_y+M_\Psi\right)f_n^\psi(y)=m_{f,n} f_n^\eta(y)\label{psi_KK_EoM} \, , 
 \end{align}
together with the boundary conditions for $f_n^\psi$,
\beq
    f_n^\psi(0) = f_n^\psi(\pi R) = 0\, . \label{psi_KK_BC}
\eeq
For the consistency between  Eq.~\eqref{eta_KK_EoM} and Eq.~\eqref{psi_KK_BC}, we have the boundary condition for $f_n^\eta(y)$,
\begin{align}
    \lmk \del_y+M_\Psi\rmk f_n^\eta|_{0,\pi R} = 0\, . \label{eta_KK_BC}
\end{align}
We can multiply $(\mp \del_y+M_\Psi)$ on the both sides of Eqs.~\eqref{eta_KK_EoM},~\eqref{psi_KK_EoM} to get the ordinary eigenvalue equations,
\beq
 (-\del_y^2 +M_\Psi^2) f_n^{\psi(\eta)} = m_{f,n}^2 f_n^{\psi(\eta)}\, .\label{fermion_KK_EoM}
\eeq
Solving Eqs.~\eqref{psi_KK_BC}--\eqref{fermion_KK_EoM}, we find the eigenmasses,
\begin{align}
    m_{f,n}^2 = M_\Psi^2 +\frac{n^2}{R^2}\, ,
\end{align}
and the KK profiles,
\begin{align}
    f_0^\psi = 0 \, &,\quad f_n^\psi = \sqrt{\frac{2}{\pi R}} \sin\frac{ny}{R} \quad \text{for} ~ n\geq 1\, , \\[1ex]
    f_0^{\eta} = N_0 e^{-M_\Psi y}\, &, \quad f_n^\eta = \sqrt{\frac{2}{\pi R}} \lmk \frac{M_\Psi}{m_n}\sin\frac{ny}{R} - \frac{n}{m_n R}\cos\frac{ny}{R} \rmk \quad \text{for} ~ n\geq 1\, , 
\end{align}
where the normalization factors of the zero mode $N_0=\left(\int_0^{\pi R}\dd y\,e^{-2M_\Psi y}\right)^{-1/2}$ and the massive modes are fixed by the canonically normalized kinetic terms,
\begin{align}
    \int_0^{\pi R} \dd y \, f_m^{\psi(\eta)}(y)f_n^{\psi(\eta)}(y)=\delta_{mn} \, .
\end{align}

For the bulk scalar, by performing the action variation upon Eq.~\eqref{5D_scalar},
\begin{align}
    \delta S_5^s &=-\int d^4x\int_0^L dy\,\left[\delta Q^*\left(D_MD^M+M_Q^2\right)Q+\mathrm{h.c.}\right]\notag  \\
    &-\int d^4x\,\left[\delta Q^* \lmk D_y Q \rmk+\mathrm{h.c.}\right]^{\pi R}_0\, ,
\end{align}
we get the EoMs and the general boundary conditions,
\begin{align}
    &\left(D_MD^M+M_Q^2\right)Q=0 \, ,\\[1ex]
    & \lkk \delta Q^* D_y Q + (D_y Q)^*\delta Q \rkk_{0,\pi R} = 0 \, ,
\end{align}
where the gauge field inside the covariant derivative can be neglected when we calculate the KK spectrum.
Decomposing $Q(x,y)=\sum_{n=0}^\infty Q_n(x)f_n^Q(y)$, we find the equations for the profiles $f_n^Q (y)$,
\begin{align}
    \left(-\partial_y^2+M_Q^2\right)f_n^Q(y)=m_{Q,n}^2 f_n^Q(y) \, ,\qquad \partial_y f_n^Q(0)=\partial_y f_n^Q(\pi R)=0 \, .
\end{align}
Here, the Neumann boundary condition is chosen as in Eq.~\eqref{scalar_BC} to make the boundary variation vanish.
Therefore, the KK spectrum of $Q(x,y)$ can be solved as
\begin{align}
	m_{Q,n}^2 = M_Q^2 + \frac{n^2}{R^2}\, ,\quad f_0^Q(y) =\sqrt{\frac{1}{\pi R}}\, ,\quad  f_n^Q(y) = \sqrt{\frac{2}{\pi R}} \cos\frac{ny}{R}\, ,\quad n>0\, , 
\end{align}
where the normalization is chosen in the same way as the fermion case so that the kinetic terms are canonically normalized.

\section{From Gauged WZW to 5D CS Term \label{sec:appendix_WZW_CS}}
The gauged WZW term has been derived in Sec.~\ref{sec:WZW}. Here, we show how it reproduces the 5D CS term in the continuum limit.
We start from the unit WZW term,
\begin{align}
    S_{{\rm WZW},i}^{\rm mix} &= \int_{R^4} \frac{1}{16\pi^2}\lkk\lmk A_i+ A_{i+1}+\tilde{\Phi}_i^* \dd \tilde{\Phi}_i\rmk \tr\lmk G_{i+1}\dd G_{i+1} +\frac{2}{3} G_{i+1}^3\rmk -\frac{1}{3}A_i\tr\lmk\tilde{\Sigma}_i\dd\tilde{\Sigma}_i^\dag\rmk^3 \right. \notag\\
	& \left. + \dd A_i \tr\lmk \tilde{\Sigma}_i^\dag \dd \tilde{\Sigma}_i G_{i+1} +G_i\tilde{\Sigma}_i G_{i+1}\tilde{\Sigma}_i^\dag +G_i\tilde{\Sigma}_i \dd\tilde{\Sigma}_i^\dag\rmk  -{\rm p.c.}\rkk  \nonumber \\
    & = \int_{R^4}\frac{1}{16\pi^2} \Bigg[  \lmk A_i+ A_{i+1}+\tilde{\Phi}_i^* \dd \tilde{\Phi}_i\rmk \tr\lmk G_{i+1}\dd G_{i+1}+\frac{2}{3} G_{i+1}^3\rmk \notag \\
    & -\lmk A_i + A_{i+1}+\tilde \Phi_i \dd \tilde \Phi_i^*\rmk \tr \lmk G_{i} \dd G_{i}+\frac{2}{3}G_{i}^3\rmk -\frac{1}{3}(A_i+A_{i+1})\lmk\tilde \Sigma_i\dd \tilde{\Sigma}_i^\dag\rmk^3 \notag \\
    &+(\dd A_i +\dd A_{i+1})\tr\lmk \tilde\Sigma_i^\dag G_i\tilde\Sigma_i G_{i+1} +G_i\tilde\Sigma_i \dd \tilde\Sigma_i^\dag +\tilde\Sigma_i^\dag \dd\tilde \Sigma_i G_{i+1}\rmk \Bigg]\, ,
\end{align}
and replace 
\begin{align}
    A_i\leftrightarrow A\, , A_{i+1}\leftrightarrow A+a\del_5 A\, , \tilde{\Phi}_i \leftrightarrow 1+aA_5\, , G_i\leftrightarrow G\, , G_{i+1}\leftrightarrow G+a\del_5 G\, ,\tilde{\Sigma}_i\leftrightarrow 1+a G_5 \, .
\end{align}
Keeping only the terms linear in the interval $a$, one obtains
\begin{align}
    &\lmk A_{i}+A_{i+1}+ \tilde\Phi_i^* \dd \tilde\Phi_i\rmk \tr \lmk G_{i+1} \dd G_{i+1}+\frac{2}{3}G_{i+1}^3 \rmk \notag\\
   &\to (2A+a\del_5 A + a\dd A_5)\tr\lmk(G+a\partial_5 G)\dd(G+a\partial_5 G)+\frac{2}{3}(G+a\partial_5 G)^3\rmk\notag\\
   &=(2A+a\del_5 A + a\dd A_5) \tr\lmk G\dd G+\frac{2}{3}G^3+a\partial_5 G \dd G+aG\partial_5 \dd G+2aG^2\partial_5 G\rmk\notag\\
   &=a(\del_5 A + \dd A_5)\tr \lmk G\dd G+\frac{2}{3}G^3\rmk
   +2aA\tr \lmk \partial_5 G\dd G+G\partial_5\dd G+2 G^2\partial_5 G\rmk \, ,
\end{align}
\begin{align}
     -\lmk A_i + A_{i+1}+\tilde \Phi_i \dd \tilde \Phi_i^*\rmk \tr \lmk G_{i} \dd G_{i}+\frac{2}{3}G_{i}^3\rmk 
     &\to -(2A+a \partial_5 A-a  \dd A_5)\tr \lmk G \dd G+\frac{2}{3}G^3\rmk \notag\\
     &=a(\dd A_5-\del_5 A)\tr\lmk G \dd G+\frac{2}{3}G^3\rmk \, ,
\end{align}
\begin{align}
    (\dd A_{i}+\dd A_{i+1})\tr(\tilde \Sigma_i^\dagger G_{i}\tilde \Sigma_i G_{i+1})
    &\to (2\dd A+a\dd \partial_5 A)\tr\lmk(1-aG_5)G(1+aG_5)(G+a\partial_5 G)\rmk \notag \\
    &= (2\dd A+a\dd \partial_5 A)\tr \lmk G^2-a G_5 G^2+a G G_5 G+a G\partial_5 G\rmk \notag\\
    &=2a\dd A \tr(-2G_5 G^2+ G\partial_5 G)\, ,
\end{align}
\begin{align}
    (\dd A_{i}+\dd A_{i+1})
    \tr( G_{i} \tilde \Sigma_i \dd\tilde\Sigma_i^\dagger)
    &\to (2\dd A + a\del_5 A)\tr\lmk G(1+aG_5)\dd (1-aG_5)\rmk \notag\\
    &=-2a\dd A \tr (G\dd G_5) \, ,
\end{align}
\begin{align}
    (\dd A_{i}+\dd A_{i+1})
   \tr(\tilde{\Sigma}_i^\dagger \dd \tilde{\Sigma}_i G_{i+1})
   &\to (2\dd A+a\dd \partial_5 A)\tr\lmk(1-aG_5)\dd(1+aG_5)(G+a\del_5 G)\rmk\notag\\
   &=2a\dd A\tr( \dd G_5 G)  \, ,
\end{align}
\begin{align}
    -\frac{1}{3}(A_{i}+A_{i+1})\tr((\tilde \Sigma_i^\dagger \dd\tilde\Sigma_i)^3)&\to 0\ .
\end{align}
Here, all the terms are differential 4-forms and $A_5, G_5$ are treated as scalars (0-forms).
The terms independent of $a$ cancel each other, which we did not show explicitly, while the higher order terms in $a$ are just neglected.
Summing over $i$ and replacing $\sum_i a\leftrightarrow \int \dd y$ lead to
\begin{align}
    S_{\rm WZW}^{\rm mix} \rightarrow \frac{1}{16\pi^2}\int \dd y\Bigg[& \int_{R^4}  2\dd A_5 \tr\lmk G \dd G+\frac{2}{3}G^3\rmk + 2A\tr \lmk \partial_5 G\dd G+G\partial_5\dd G+2 G^2\partial_5 G\rmk  \notag \\
    &+2\dd A\lmk -2G_5G^2+G\del_5 G-2G\dd G_5\rmk\Bigg] \, ,
\end{align}
where, inside the square bracket, we still take the 4-form notation and treat the fifth components as scalars.
For the first and the last terms, one can integrate by parts,
\begin{align}
    \dd A_5 \tr\lmk G \dd G+\frac{2}{3}G^3\rmk &= -A_5 \dd \tr\lmk G \dd G+\frac{2}{3}G^3\rmk =A_5\tr\lmk\mathcal{G}^2\rmk \, , \\[1ex]
    \dd A\lmk -2G_5G^2+G\del_5 G-2G\dd G_5\rmk &= A\tr(-2G^2\dd G_5 - 2GG_5\dd G+ 2G_5 G\dd G  \notag \\
    &-2\dd G\dd G_5+ \dd G\del_5 G  -G\dd\del_5 G) \, ,
\end{align}
where we use $\dd \tr\lmk G \dd G+\frac{2}{3}G^3\rmk = \tr\lmk\mathcal{G}^2\rmk$.
Altogether, we obtain
\begin{align}
    S_{\rm WZW}^{\rm mix} &\rightarrow -\frac{1}{8\pi^2}\int \dd y \Bigg[\int_{R^4} A_5\tr\lmk\mathcal{G}^2\rmk - 2A\tr\bigg( \del_5 G\dd G-\dd G\dd G_5 \notag \\
    &\hspace{3cm} +G^2\del_5 G-G^2\dd G_5 - G G_5\dd G + G_5 G\dd G\bigg)\Bigg] \, . \label{WZW_to_CS_0}
\end{align}
Next, to show the right-hand side of Eq.~\eqref{WZW_to_CS_0} is exactly the same as the 5D CS term, we work backwards and expand the CS term as
\begin{align}
    S_5^{\rm CS} & = -\int \dd^5 x\frac{\kappa_{\rm CS}}{32\pi^2}\epsilon^{MNPQR}A_M\tr \lmk \mathcal{G}_{NP}\mathcal{G}_{QR}\rmk \notag \\
    & = - \frac{\kappa_{\rm CS}}{32\pi^2}\int\dd^5 x \lkk \epsilon^{\mu\nu\rho\sigma}A_5\tr\lmk\mathcal{G}_{\mu\nu}\mathcal{G}_{\rho\sigma}\rmk - 4\epsilon^{\mu \nu\rho\sigma}A_\mu\tr\lmk \mathcal{G}_{5\nu}\mathcal{G}_{\rho\sigma}\rmk \rkk \, ,\label{WZW_to_CS_1}
\end{align}
where we separate the possible positions of the fifth-dimensional index in the 5D Levi-Civita tensor and the last four possibilities are combined into the second term.
The correspondence between the first terms in Eqs.~\eqref{WZW_to_CS_0}, \eqref{WZW_to_CS_1} is easily proved according to the definition $\mathcal{G}=\frac{1}{2}\mathcal{G}_{\mu\nu}\dd x^\mu \dd x^\nu$.
Let us further expand the trace part in the second term,
\begin{align}
    \epsilon^{\mu\nu\rho\sigma}\tr\lmk \mathcal{G}_{5\nu}\mathcal{G}_{\rho\sigma} \rmk &= \epsilon^{\mu\nu\rho\sigma}\tr\lkk(\del_5 G_\nu-\del_\nu G_5 + G_5 G_\nu-G_\nu G_5)(\del_\rho G_\sigma-\del_\sigma G_\rho+G_\rho G_\sigma-G_\sigma G_\rho)\rkk \notag \\
    & = 2\epsilon^{\mu\nu\rho\sigma}\tr [ \del_5 G_\nu \del_\rho G_\sigma -\del_\nu G_5\del\rho G_\sigma +\del_5 G_\nu G_\rho G_\sigma \notag \\
    &-\del_\nu G_5 G_\rho G_\sigma -G_\nu G_5\del_\rho G_\sigma +G_5 G_\nu\del_\rho G_\sigma ] \, .
\end{align}
Here, the simplification comes from the fact that both $\epsilon^{\mu\nu\rho\sigma}$ and $\del_\rho G_\sigma-\del_\sigma G_\rho = [G_\rho, G_\sigma]$ are anti-symmetric in $\rho,\sigma$.
The result is organized in correspondence with the second term in Eq.~\eqref{WZW_to_CS_0}, and it is not hard to see that they are the same.
Now we have proved that the gauged WZW term $S_{\rm WZW}^{\rm mix}$ reproduces in the continuum limit the 5D CS term $S_5^{\rm CS}$ with $\kappa_{\rm CS}=1$.

\section{$U(1)$ Spectrum with Alternative Dirichlet BC \label{sec:appendix_precise_spectrum}}
According to Eq.~\eqref{modified_mass_term}, the new mass-squared matrix after the two boundary gauge fields $A_{1,\mu}, A_{N,\mu}$ and Higgs fields $H_1,H_N$ are introduced is given by
\begin{align}
    {\mathcal{M}_{U(1)}'}^2 = g^2 v^2\begin{pmatrix}
1+\beta & -1 & 0 & \cdots & 0\\
-1 & 2 & -1 & \cdots & 0\\
0 & -1 & 2 & \ddots & \vdots\\
\vdots & & \ddots & \ddots & -1\\
0 & 0 & \cdots & -1 & 1+\beta
\end{pmatrix} = g^2v^2 {\Delta_N^{\rm N}}' \, ,
\end{align}
where $\beta = \lmk \frac{v'}{v}\rmk^2$ and the modified discrete Laplacian with Neumann boundary conditions is defined by
\begin{align}
    {\Delta_N^{\rm N}}'\equiv \Delta_N^{\rm N} + \beta (E_{11}+E_{NN}) \, ,
\end{align}
with the single-element matrices $(E_{11})_{ij} = \delta_{1i}\delta_{1j}$, $(E_{NN})_{ij} = \delta_{Ni}\delta_{Nj}$.
Assuming that the eigenvector of ${\Delta_N^{\rm N}}'$ is $B=(b_1,b_2,\cdots,b_N)^T$ and the eigenvalue is $\lambda$, the eigenvalue equation ${\Delta_N^{\rm N}}' B=\lambda B$ gives 
\begin{align}
    (1+\beta)b_1 -b_2 &= \lambda b_1 \, , \\[1ex]
    -b_{i-1} + 2b_i - b_{i+1} &= \lambda b_i \, ,\quad i=2,\cdots,N-1 \, , \\[1ex]
    -b_{N-1} + (1+\beta) b_N &= \lambda b_N \, .
\end{align}
One trick is to formally introduce $b_0, b_{N+1}$ and turn all the equations into the same form,
\begin{align}
     -b_{i-1} + 2b_i - b_{i+1} &= \lambda b_i \, ,\qquad i=1,\cdots,N \, , \label{interior_eq}
\end{align}
together with the constraint,
\begin{align}
    b_0 = (1-\beta) b_1 \, ,\quad b_{N+1} = (1-\beta) b_N\, . \label{endpoint_eq}
\end{align}
It is easy to check the equivalence by substituting the constraint into Eq.~\eqref{interior_eq}.
Now the system is composed of a set of bulk equations~\eqref{interior_eq} and two boundary conditions~\eqref{endpoint_eq}.
Just as in the continuum case, we take an ansatz for the general solution,
\begin{align}
    b_j = A e^{iq j} + B e^{-iqj}\, , \quad j=0,\cdots,N+1\, ,
    \label{eq:bj}
\end{align}
and Eq.~\eqref{interior_eq} gives
\begin{align}
    \mathrm{l.h.s.} &= Ae^{iqj}(-e^{-iq} +2 -e^{iq} ) + B e^{-iqj}(-e^{iq} +2 -e^{-iq} ) = b_j (2-2\cos q) \, , \notag \\[1ex]
    \mathrm{r.h.s.} &= \lambda b_j \, .
\end{align}
Therefore, we find the eigenvalue written in terms of the parameter $q$ as
\begin{align}
    \lambda = 2-2\cos q \, .
\end{align}
To finally determine the allowed value of $q$, we substitute the ansatz into the boundary conditions~\eqref{endpoint_eq}, which gives
\begin{align}
    b_j &= C \lkk \sin(jq) + (\beta -1) \sin((j-1)q)\rkk \, , \\[1ex]
    \sin[(N-1)q] + &\frac{2}{\beta-1}\sin(Nq) + \frac{1}{(\beta-1)^2}\sin[(N+1)q] = 0\, , \label{q_value_eq}
\end{align}
where $C$ is an overall factor determined by the normalization. The first equality includes how the coefficient $B$ depends on $A$ while the second equality determines the value of $q$.
Eq.~\eqref{q_value_eq} can be solved perturbatively in $\beta^{-1}$ since $\beta\gg 1$.
At the leading order, we can neglect the last two terms and obtain
\begin{align}
    \sin[(N-1)q] = 0\, ,
\end{align}
which leads to
\begin{align}
    q_k =\frac{k\pi}{N-1}\, , \quad \text{with} ~~ k=1,\cdots, N-2 \, ,
\end{align}
Note that $q$ is confined to the range $[0,\pi]$ for the independence of eigenvectors and $q=0,\pi$ lead to the trivial eigenvectors, and hence $k$ has a finite range.
The corresponding eigenvalues and eigenvectors are given by
\begin{align}
    \lambda_k = 2-2\cos\lmk \frac{k\pi}{N-1}\rmk\, , \quad 
    b_j^{(k)} = -C\beta \sin\lmk \frac{(j-1)k\pi}{N-1}\rmk \, ,
\end{align}
where we only keep the term linear in $\beta$ in the eigenvector. The original mass spectrum~\eqref{eigenmass_U1}, \eqref{eigenstate_U1_canonical} are reproduced up to the normalization factor of the eigenvectors.
We have not captured the two heavy modes yet.
The reason is that until now we have been restricted in the cases where $q$ is a real number and consequently, $\lambda=2-2\cos q\leq 4$.
However, we expect the eigenvalues of the two heavy modes to be much larger.
Now we consider $q$ as a complex number of the form,
\begin{align}
    q = \pi + i\kappa\, ,
\end{align}
with $\kappa$ being real and positive.
Since $\cos(\pi + i\kappa) = -\cosh\kappa$ and $\sin[N(\pi + i\kappa)]= (-1)^N i\sinh(N\kappa)$, the eigenvalue and Eq.~\eqref{q_value_eq} are changed to
\begin{align}
    \lambda_{\rm heavy} &= 2+2\cosh\kappa \, , \\[1ex]
    \sinh[(N+1)\kappa] - 2(\beta-1)\sinh(N\kappa)& + (\beta-1)^2\sinh[(N-1)\kappa] = 0 \, .
\end{align}
Naturally we have $N\kappa\gg 1$ for a large lattice and thus $\sinh(N\kappa)\simeq \half e^{N\kappa}$.
Then one gets a simplified equation for $\kappa$,
\begin{align}
    e^{(N+1)\kappa} -2(\beta-1)e^{N\kappa} +(\beta -1)^2 e^{(N-1)\kappa}\simeq 0\, .
\end{align}
Factoring out $e^{(N-1)\kappa}$, we obtain
\begin{align}
    e^{2\kappa} -2(\beta -1) e^\kappa +(\beta -1)^2 \simeq 0\, ,
\end{align}
and therefore,
\begin{align}
    \kappa\simeq \ln(\beta-1)\, .
\end{align}
The resulting eigenmass squared is given by
\begin{align}
    m_{\rm heavy}^2 = g^2 v^2\lambda_{\rm heavy} \simeq g^2v^2\lmk 1 + \beta +\frac{1}{\beta-1}\rmk \, ,
\end{align}
which just deviates from our naive estimation in the decoupling limit by a sub-leading term.
Substituting $q = \pi + i\kappa$ into the general solution~\eqref{eq:bj}, we can see that two modes have peaks at two endpoints separately, 
\begin{align}
    b_j \sim A (-1)^j e^{-\kappa j} - Be^{\kappa N}  (-1)^j e^{- \kappa (N-j)}\, ,
\end{align}
which correspond to $\hat A_{1,\mu}$ and $\hat A_{N,\mu}$.

\bibliographystyle{JHEP}
\bibliography{reference}

\end{document}